\documentclass[a4paper,fleqn]{cas-dc}
\usepackage{float}
\usepackage{subfigure}
\usepackage[ruled]{algorithm2e}
\usepackage{bm}
\usepackage{amsmath} 
\usepackage{amssymb}
\usepackage{makecell}
\usepackage{bbding}
\usepackage{diagbox}

\usepackage{changes}
\usepackage{xcolor}

\usepackage{listings}
\usepackage{color}
\usepackage{hyperref}
\definecolor{dkgreen}{rgb}{0,0.6,0}
\definecolor{gray}{rgb}{0.5,0.5,0.5}
\definecolor{mauve}{rgb}{0.58,0,0.82}
\usepackage[authoryear]{natbib}

\def\tsc#1{\csdef{#1}{\textsc{\lowercase{#1}}\xspace}}
\tsc{WGM}
\tsc{QE}

\begin{document}
\let\WriteBookmarks\relax
\def\floatpagepagefraction{1}
\def\textpagefraction{.001}
    
\shorttitle{ADS\_UNet: A Nested UNet for Histopathology Image Segmentation}    

\shortauthors{Yang, Dasmahapatra, Mahmoodi}  

\title [mode = title]{ADS\_UNet: A Nested UNet for Histopathology Image Segmentation}  
\author[1]{Yilong Yang}[orcid=0000-0002-2595-7883]
\cormark[1]


\ead{Yilong.Yang@soton.ac.uk}
\credit{Conceptualization, Methodology, Visualization, Investigation, Software, Formal analysis, Writing-Original Draft}

\affiliation[1]{organization={School of Electronics and Computer Science, University of Southampton},
            addressline={University Road}, 
            city={Southampton},
            postcode={SO17 1BJ}, 
            state={Hampshire},
            country={United Kingdom}}


\author[1]{Srinandan Dasmahapatra}
\ead{sd@ecs.soton.ac.uk}
\credit{Supervision, Conceptualisation, Formal analysis, Writing-Reviewing \& Editing}

\author[1]{Sasan Mahmoodi}
\ead{sm3@ecs.soton.ac.uk}
\credit{Supervision, Conceptualisation, Formal analysis, Writing-Reviewing \& Editing}


\cortext[1]{Corresponding author}



\begin{abstract}
The UNet model consists of fully convolutional network (FCN) layers arranged as contracting encoder and upsampling decoder maps. Nested arrangements of these encoder and decoder maps give rise to extensions of the UNet model, such as UNete and UNet++. Other refinements include constraining the outputs of the convolutional layers to discriminate between segment labels when trained end to end, a property called deep supervision. This reduces feature diversity in these nested UNet models despite their large parameter space. Furthermore, for texture segmentation, pixel correlations at multiple scales contribute to the classification task; hence, explicit deep supervision of shallower layers is likely to enhance performance. In this paper, we propose ADS UNet, a stage-wise additive training algorithm that incorporates resource-efficient deep supervision in shallower layers and takes performance-weighted combinations of the sub-UNets to create the segmentation model. We provide empirical evidence on three histopathology datasets to support the claim that the proposed ADS UNet reduces correlations between constituent features and improves performance while being more resource efficient. We demonstrate that ADS\_UNet outperforms state-of-the-art Transformer-based models by 1.08 and 0.6 points on CRAG and BCSS datasets, and yet requires only 37\% of GPU consumption and 34\% of training time as that required by Transformers.
\end{abstract}

\begin{highlights}
\item We propose ADS\_UNet that integrates cascade training and AdaBoost algorithm.
\item We supervise layers of UNet to learn useful features in a manner that is learnable.
\item The importance of layers varies across the training and contributes differently.
\item The ADS\_UNet achieves state-of-the-art performance.
\item The ADS\_UNet is much more memory and computationally efficient than Transformers.
\end{highlights}

\maketitle
\section{Introduction}
The fully convolutional neural network (FCN) \citep{long2015fully}, trained end-to-end on per-pixel labels, is considered a milestone in image segmentation using deep networks. It was then extended by \cite{ronneberger2015u} to include a large number of up-sampled features concatenated using skip connections with the encoded convolutional features. They named the network a UNet after a geometrical laying out of the network topology in a u-shape. \cite{zhou2019unet++} modified the UNet architecture by adding more nodes and connections to capture low-level correlation of distributed semantic attributes. The resulting architectures, known as UNet$^e$ ($e$ denotes ensemble) and UNet++, used class labels to guide the outputs of decoder layers (called deep supervision) to learn highly discriminative features. 

Both UNet$^e$ and UNet++ can be classified as ensemble models, in which multiple models are created to obtain better performance than each constituent model alone \citep{opitz1999popular}. A property that is present in a good ensemble is the diversity of the predictions made by contributing models. However, end-to-end training of deep networks tends to correlate intermediate layers \citep{ji2020directional}, hence the collaborative learning of constituent UNets adopted by UNet$^e$ and UNet++ induces learned features to be correlated. Such learning runs counter to the idea of feature diversity pursued by ensemble models. Moreover, simple averaging performed in UNet$^e$, disregarding the difference in the performance of each member also restricts the final predictive performance of the ensemble.


Based on the work of UNet$^e$ and UNet++, we pose several questions: 1) can each constituent model be forced to extract decorrelated features during training, to guarantee prediction diversity? 2)
can the outputs of constituent models, sensitive to different spatial resolutions, be weighted differently when they are integrated into the final segmentation? 3) can we provide deep supervision for encoders directly rather than by supervising  the up-sampled decoders? To address these questions, we propose the Adaboosted Deeply Supervised UNet (ADS\_UNet). The key contributions of our work can be summarized as follows:
\begin{itemize}
\item[1)]
We integrate deep supervision, cascade learning, and AdaBoost into the proposed ADS\_UNet, a stage-wise additive training algorithm, in which multiple UNets of varying depths are trained sequentially to enhance the feature diversity of constituent models. Extensive experiments demonstrate that ADS\_UNet is effective in boosting segmentation performance.
\item[2)]
In our deep supervision scheme, we down-sample the mask to have the same size as feature maps of hidden layers to compute pixel-wise loss, instead of up-sampling features. This modification retains the advantages of deep supervision and yet reduces computation cost and GPU memory consumption.
\item[3)] Instead of assigning balanced weights to all supervised layers, we introduce a learnable weight for the loss of each supervised layer to characterize the importance of features learned by layers.
\item[4)]
We conduct a comprehensive ablation study to systematically analyze the performance gain achieved by the ADS\_UNet.
\end{itemize}
\begin{figure*}[!t]
	\centering
	\subfigure[UNet$^e$]{
		\label{fig:Unet_e}
		\includegraphics[scale=0.23]{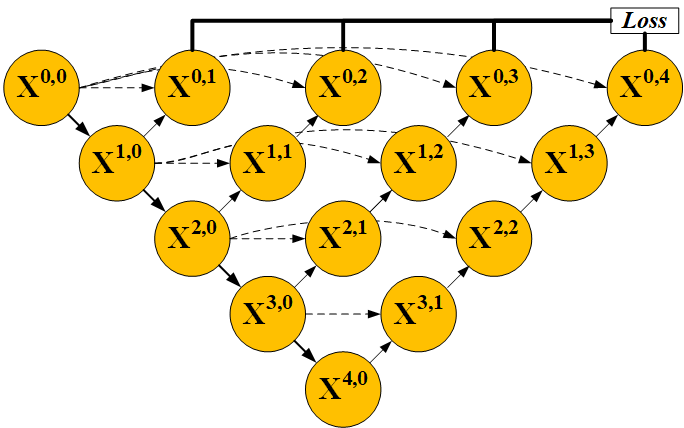}}
	\subfigure[UNet++]{
		\label{fig:UNet++}
		\includegraphics[scale=0.23]{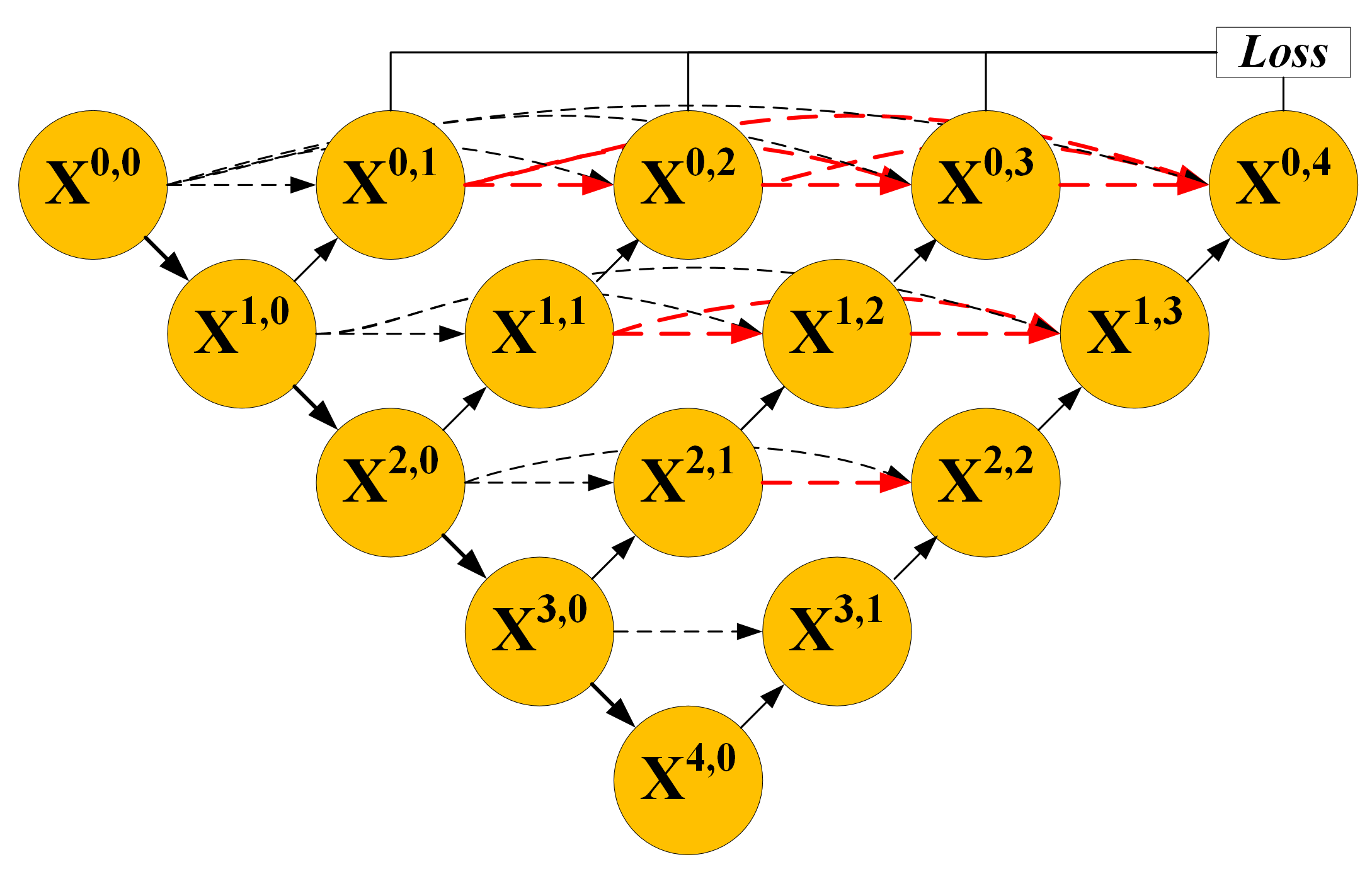}}
	\subfigure[CENet]{
		\label{fig:CENet}
		\includegraphics[scale=0.23]{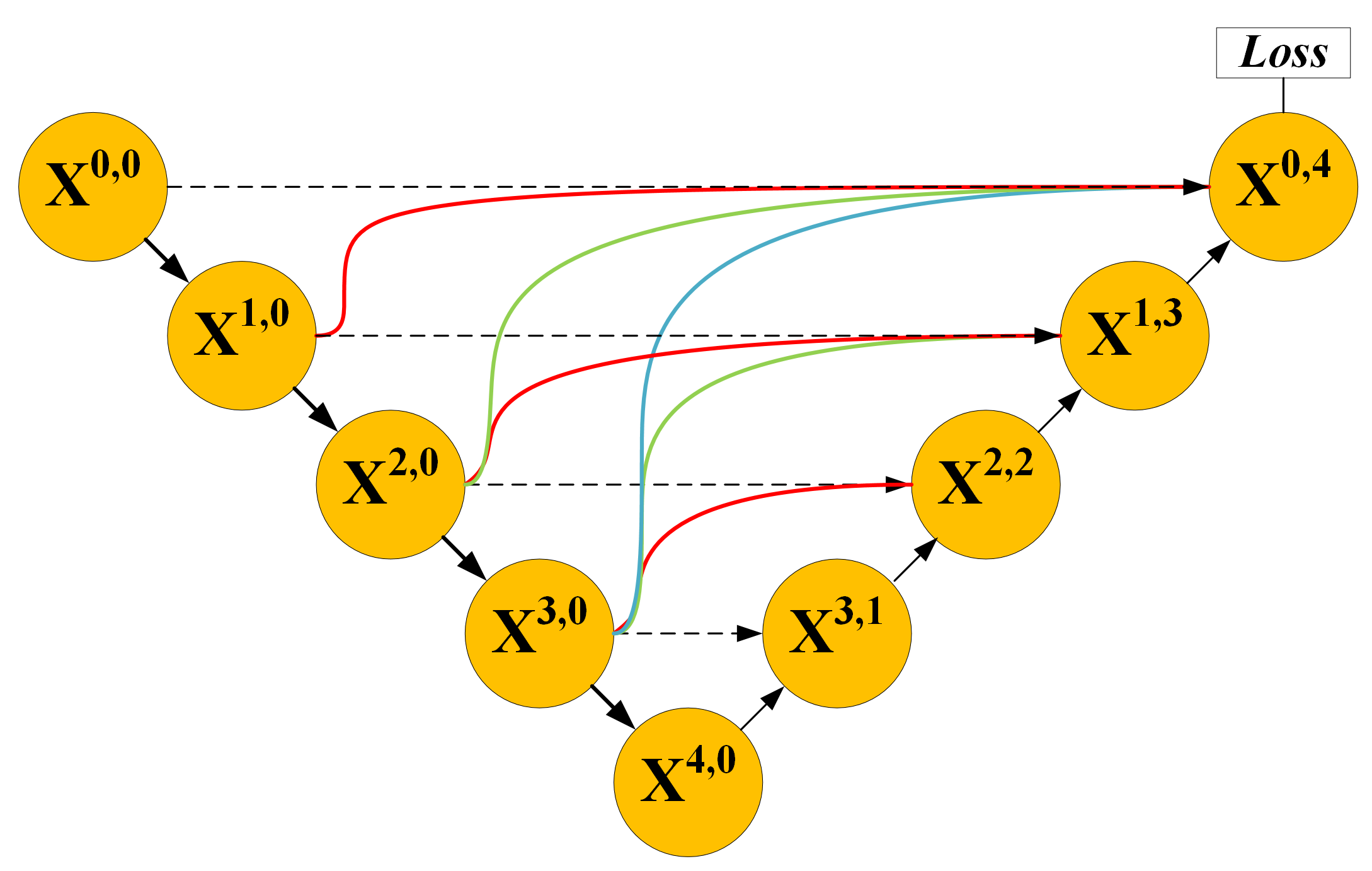}}
	\subfigure[Legend]{
  \label{fig:DSU-Net}
  \includegraphics[scale=0.35]{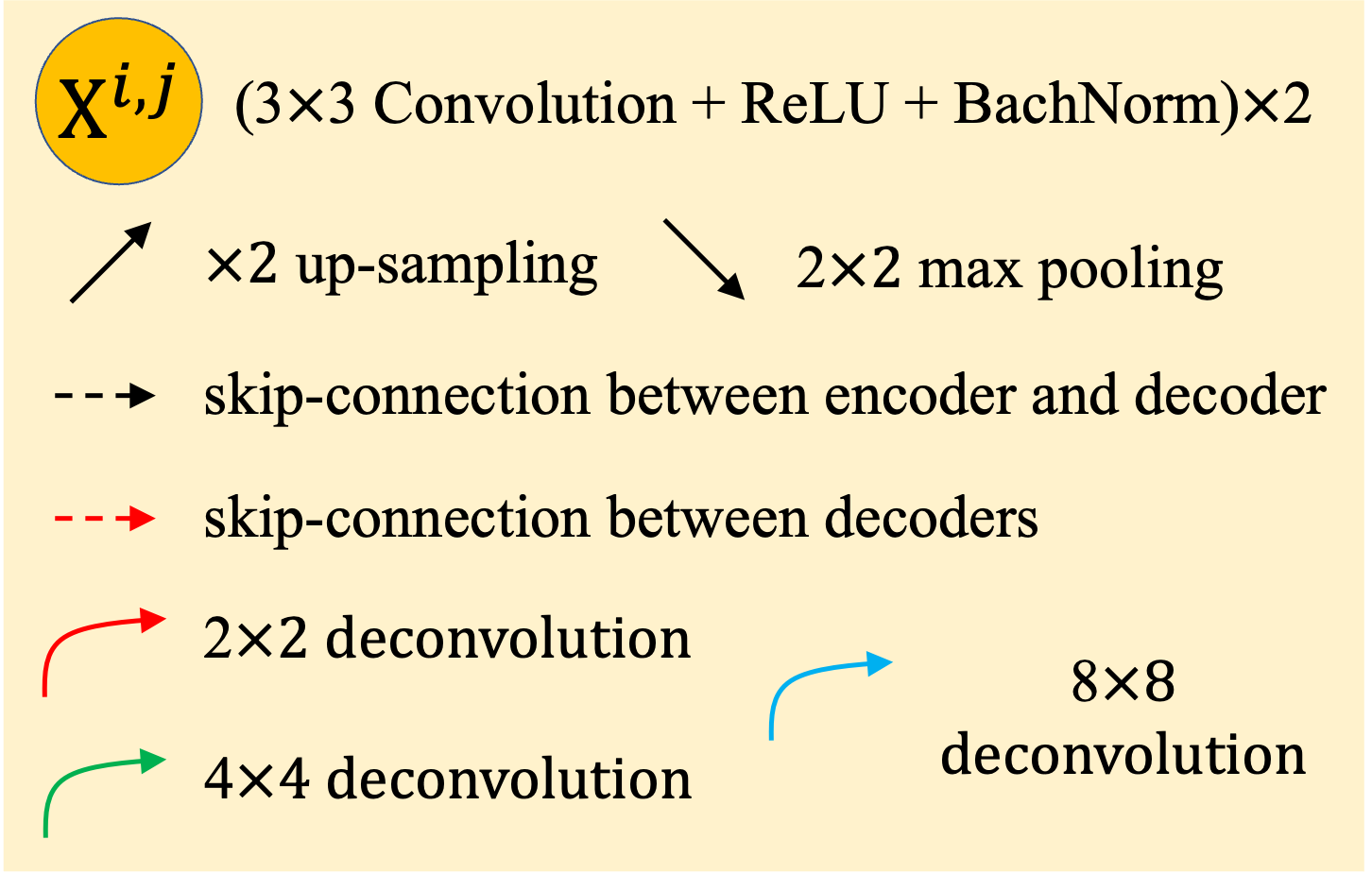}}
	\caption{Comparison of UNet$^e$ (a), UNet++ (b), CENet (c), and proposed UNet (c). UNet++ is constructed from UNet$^e$ by introducing skip-connections (red dashed lines in (b)) between decoder nodes. CENet disregards inner decoder nodes and adopts deconvolution and concatenation to harvest multi-scale context clues between encoder and decoder nodes.}
	\label{fig:compare_unet}
\end{figure*}
\section{Related Work}
In this section, we review the works related to UNet and its variants, deep supervision and AdaBoost, which are the main components of our architecture.
\subsection{UNet family}
UNet \citep{ronneberger2015u} consists of a down-sampling path to capture context, and a symmetric up-sampling path to expand feature maps back to the input size. The down-sampling part has an FCN-like architecture that extracts features with 3$\times$3 convolutions. The up-sampling part uses deconvolution to reduce the number of feature maps while increasing their area. Feature maps from the down-sampling part of the network are copied and concatenated to the up-sampling part to ensure precise localization. 

Building on the success of UNet, several variants have been proposed to further improve segmentation performance. Here we describe the networks UNet$^e$ and UNet++ \citep{zhou2019unet++}, whose simplified architectures are given in Figure~\ref{fig:compare_unet}. UNet$^e$ is an ensemble architecture, which combines UNets of varying depths into one unified structure. Note that deep supervision is required to train UNet$^e$ in an end-to-end fashion. In order to allow deeper UNets to offer a supervision signal to the decoders of the shallower UNets in the ensemble and address the potential loss of information, the UNet++ connects the decoder nodes, to enable dense feature propagation along skip connections and thus more flexible feature fusion at the decoder nodes. The difference between UNet++ and UNet$^e$ is that there are skip-connections between decoder nodes in UNet++ (highlighted in red in Figure~\ref{fig:UNet++}). \cite{zhou2022contextual} proposed the contextual ensemble network (CENet), where the contextual cues are aggregated via densely up-sampling the features of the encoder layers to the features of the decoder layers. This enables CENet to capture multi-scale context information. While UNet++ and CENet yield higher performance than UNet, it does so by introducing dense skip connections that result in a huge increase of parameters and computational cost.

Most recently, building upon the success of Vision Transformer \citep{DBLP:conf/iclr/DosovitskiyB0WZ21} on image classification tasks, self-attention modules have also been integrated into UNet-like architectures for accurate segmentation. \cite{luo2021hybrid} proposed the hybrid ladder transformer (HyLT), in which the authors use bidirectional cross-attention bridges at multiple resolutions for the exchange of local and global features between the CNN- and transformer-encoding paths. The fusion of local and global features renders HyLT robust compared to other CNN-, transformer- and hybrid- methods for image perturbations. \cite{gao2022data} presented MedFormer, in which an efficient bidirectional multi-head attention (B-MHA) is proposed to eliminate redundant tokens and reduce the quadratic complexity of conventional self-attention to a linear level. Furthermore, the B-MHA liberates the constraints of model design and enables MedFormer to extract global relations on high-resolution token maps towards the fine-grained boundary modelling. \cite{ma2022ht} proposed a hierarchical context-attention transformer-based architecture (HT-Net), which introduces an axial attention layer to model pixel dependencies of multi-scale feature maps, followed by a context-attention module that captures context information from adjacent encoder layers.
\printcredits
\subsection{Deep supervision}
A deeply supervised network (DSN) \citep{lee2015deeply} introduced classification outputs to hidden layers as well as the last layer output as is the convention. This was shown to increase the discriminative power of learned features in shallow layers and robustness to hyper-parameter choice.

Despite the fact that the original DSN was proposed for classification tasks, deep supervision can also be used for image segmentation. \cite{dou20163d} introduced deep supervision to combat potential optimization difficulties and concluded that the model acquired a faster convergence rate and greater discriminability. Based on the UNet architecture, \cite{zhu2017deeply} introduced a supervision layer to each encoder/decoder block. Their method is very similar to our proposed supervision scheme; the difference lies in how the loss between the larger-sized ground-truth and the smaller-sized output of hidden layers is calculated. Note that the dimension of feature maps of the hidden layers are gradually reduced and become much smaller than that of the ground-truth mask, because of the down-sampling operation. In \cite{dou20163d} and \cite{zhu2017deeply}, deconvolutional layers were used to up-sample feature maps back to the same size as the ground-truth mask. Evidently, the additional deconvolutional layers introduce more parameters and more computational overhead. Although it was pointed out in \cite{long2015fully} that one can learn arbitrary interpolation functions, bilinear interpolation was adopted in \cite{xie2015holistically} to up-sample feature maps with no reduction in performance compared to learned deconvolutions. All of the aforementioned literature solve the dimension mismatch problem by up-sampling feature maps. However, in our deep supervision scheme, we perform average pooling to down-sample the ground-truth mask to the same size as feature maps of hidden layers. This reduces the amount of computation and is more GPU memory efficient.
\subsection{AdaBoost}
AdaBoost (Adaptive Boosting) \citep{freund1997decision} is a very successful ensemble classifier, which has been widely used in binary classification tasks. The idea of AdaBoost is based on the assumption that a highly accurate prediction rule can be obtained by combining many relatively weak and inaccurate rules. This was re-derived in \cite{friedman2001greedy} as a gradient of an exponential loss function of a stage-wise additive model. Such an additive model was extended to the multi-class case by \cite{hastie2009multi}, who proposed SAMME (Stage-wise Additive Modeling using a Multi-class Exponential loss function) that naturally extends the original AdaBoost algorithm to the multi-class case without reducing it to multiple two-class problems. The detailed iterative procedure of multi-class AdaBoost is described in Algorithm 2 of \cite{hastie2009multi}.

Starting from equally weighted training samples, the AdaBoost trains a classifier $f_t$, ($t \in \{1,2,...,T\}$ is the iteration index) iteratively, re-weighting the training samples in the process. A misclassified item $x_i$ is assigned a higher weight $w_i^t$ so that the next iteration of the training pays more attention to it. After each classifier $f_t$ is trained, it is assigned a weight based on its error rate $\epsilon_t$ on the training set. For the integrated output of the classifier ensemble, the more accurate classifier is assigned a larger weight $\alpha_t$ to have more impact on the final outcome. A classifier with  $<\frac{1}{C}$\% accuracy (less than random guessing for $C$ target classes) is discarded. $T$ classifiers will be trained after T iterations of this procedure. The final labels can be obtained by the weighted majority voting of these $T$ classifiers.

 An adaptive algorithm, Adaboost-CNN, which combines multiple CNN models for multi-classification was introduced in \cite{taherkhani2020adaboost}. In AdaBoost-CNN, all the weak classifiers are convolutional neural networks and have the same architecture. Instead of training a new CNN from scratch, they transfer the parameters of the prior CNN to the later one and then train the new CNN for only one epoch. This achieves better performance than the single CNN, but at the cost of increasing the number of parameters several fold. Curriculum learning \citep{bengio2009curriculum} is related to boosting algorithms, in that the training schedule gradually emphasizes the difficult examples. \cite{cui2022deep} demonstrated that better performance can be achieved by forcing UNet to learn from easy to difficult scenes. However, the difficulty level of training samples is predefined according to the size of the target to be segmented, rather than calculated by the network itself.
 
\section{Method}\label{section:method}
Ensemble learning is often justified by the heuristic that each base learner might perform well on some data and less accurately on others for some learned features, to enable the ensemble to override common weaknesses. To this end, we seek enhanced segmentation performance of the model by enabling diverse feature maps to be learned. We propose the ADS\_UNet algorithm, which adopts a layer-wise cascade training approach \citep{fahlman1989cascade, bengio2007greedy, marquez2018deep} but with an added component that re-weights training samples to train each base learner in sequence. We evaluate the role of feature map diversity in section~\ref{section:Feature_similarity}.
\begin{figure*}[!t]
	\centering
	\subfigure[UNet$^1$]{
		\label{U1}	\includegraphics[scale=0.38]{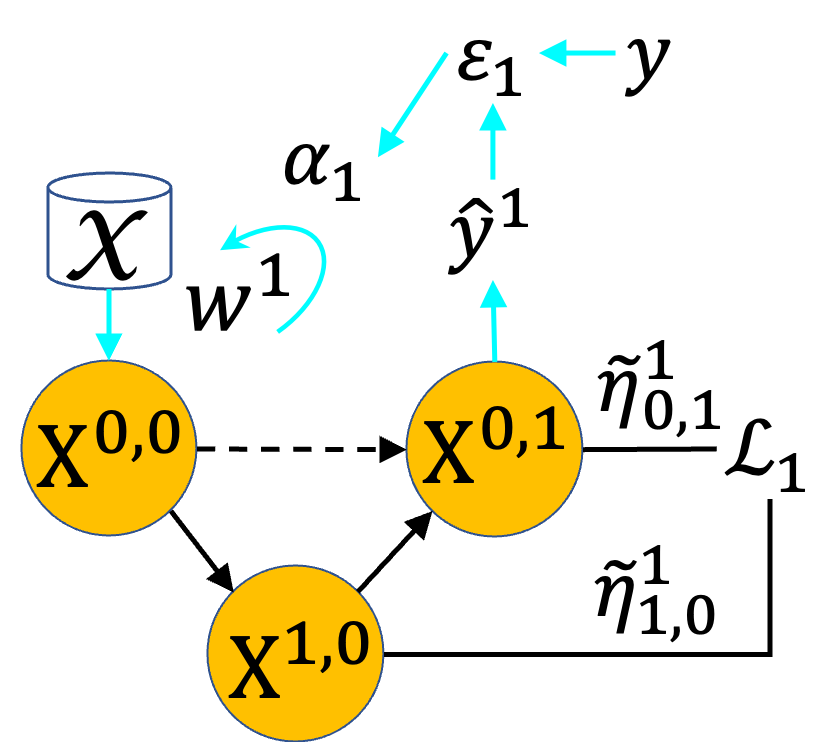}}
	\subfigure[UNet$^2$]{
		\label{U2}	\includegraphics[scale=0.38]{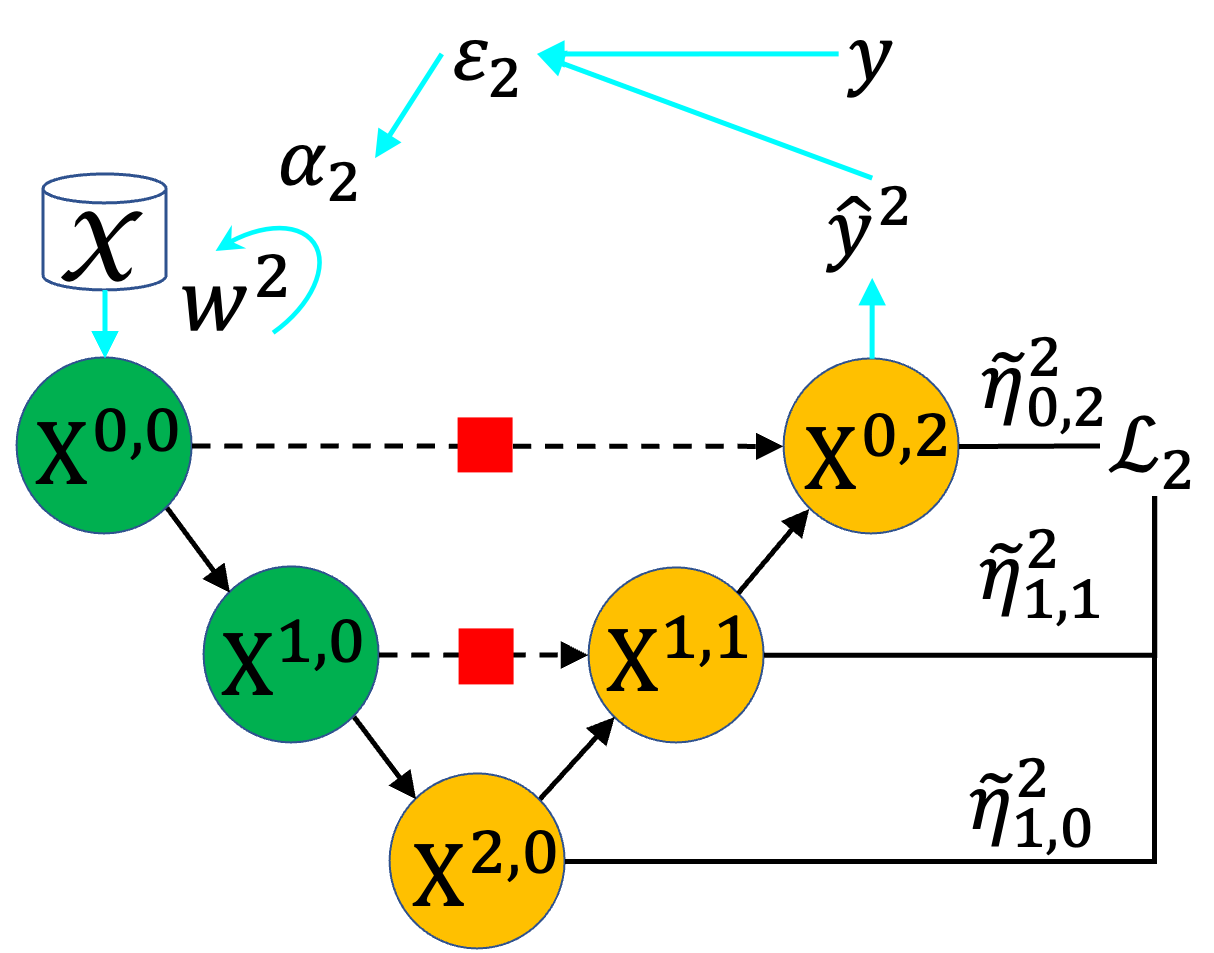}}
	\subfigure[UNet$^3$]{
		\label{U3}
		\includegraphics[scale=0.38]{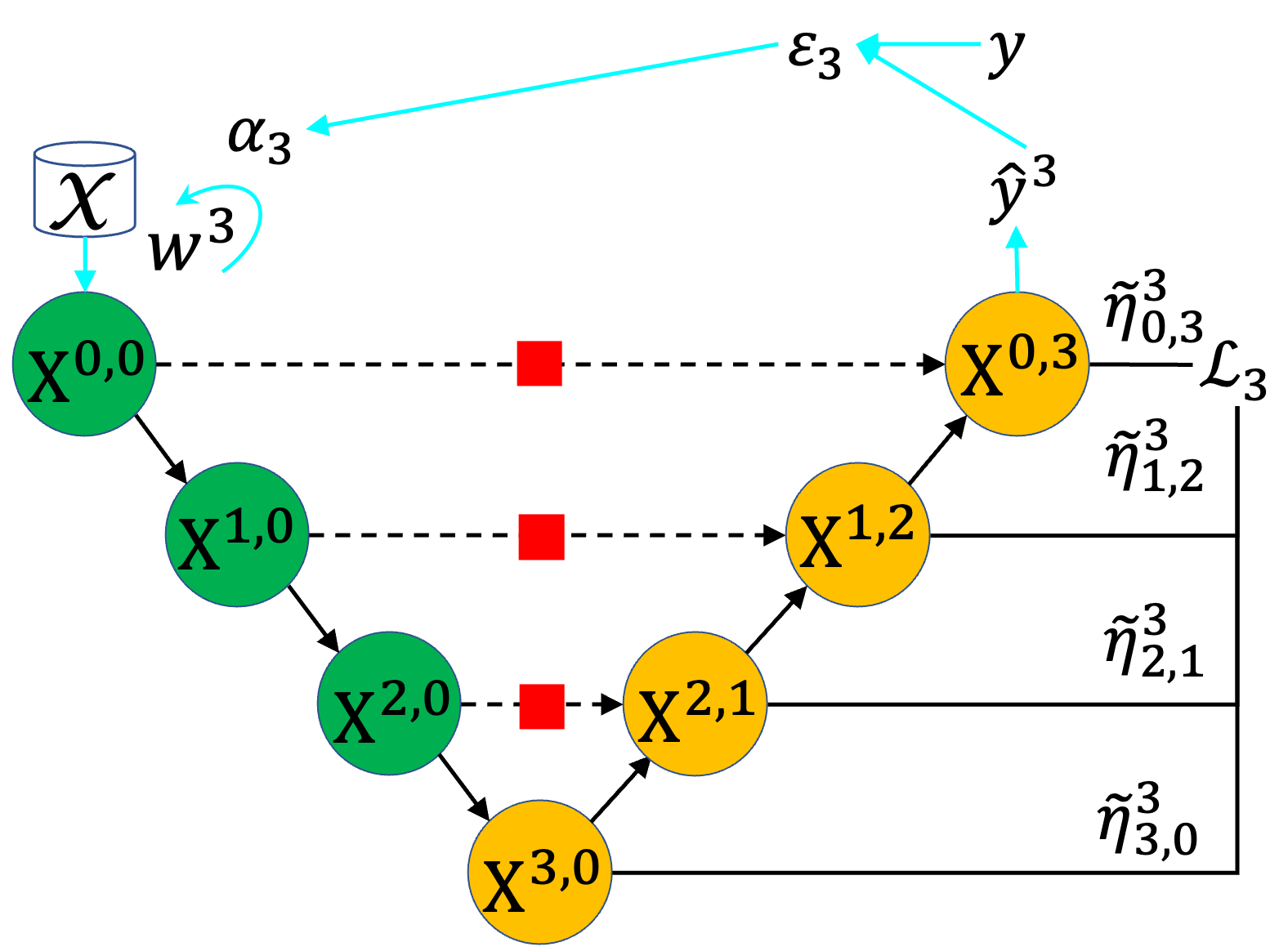}}
	\subfigure[UNet$^4$]{
		\label{U4}
		\includegraphics[scale=0.38]{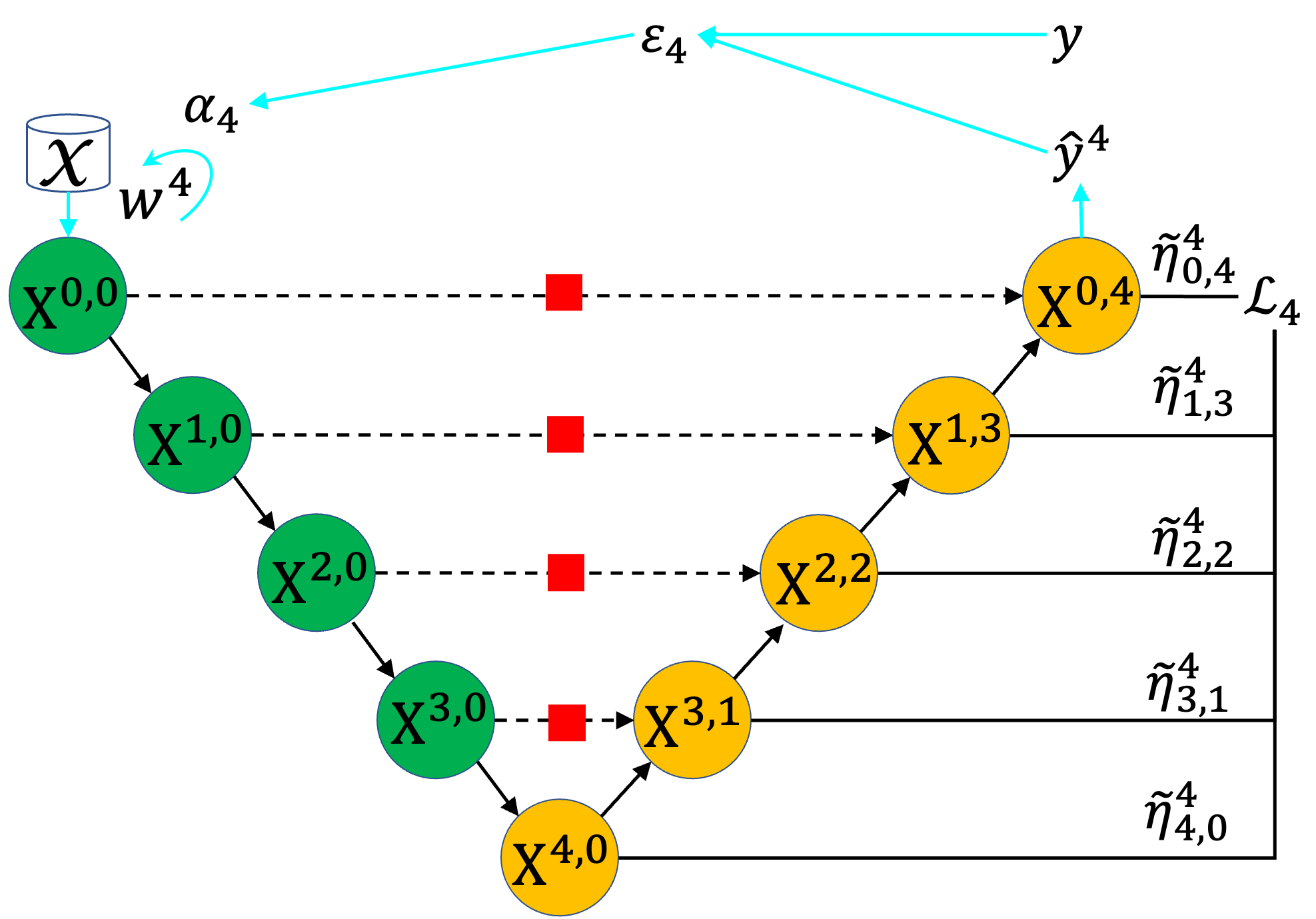}}
	\subfigure[ADS\_UNet]{
		\label{U5}
		\includegraphics[scale=0.38]{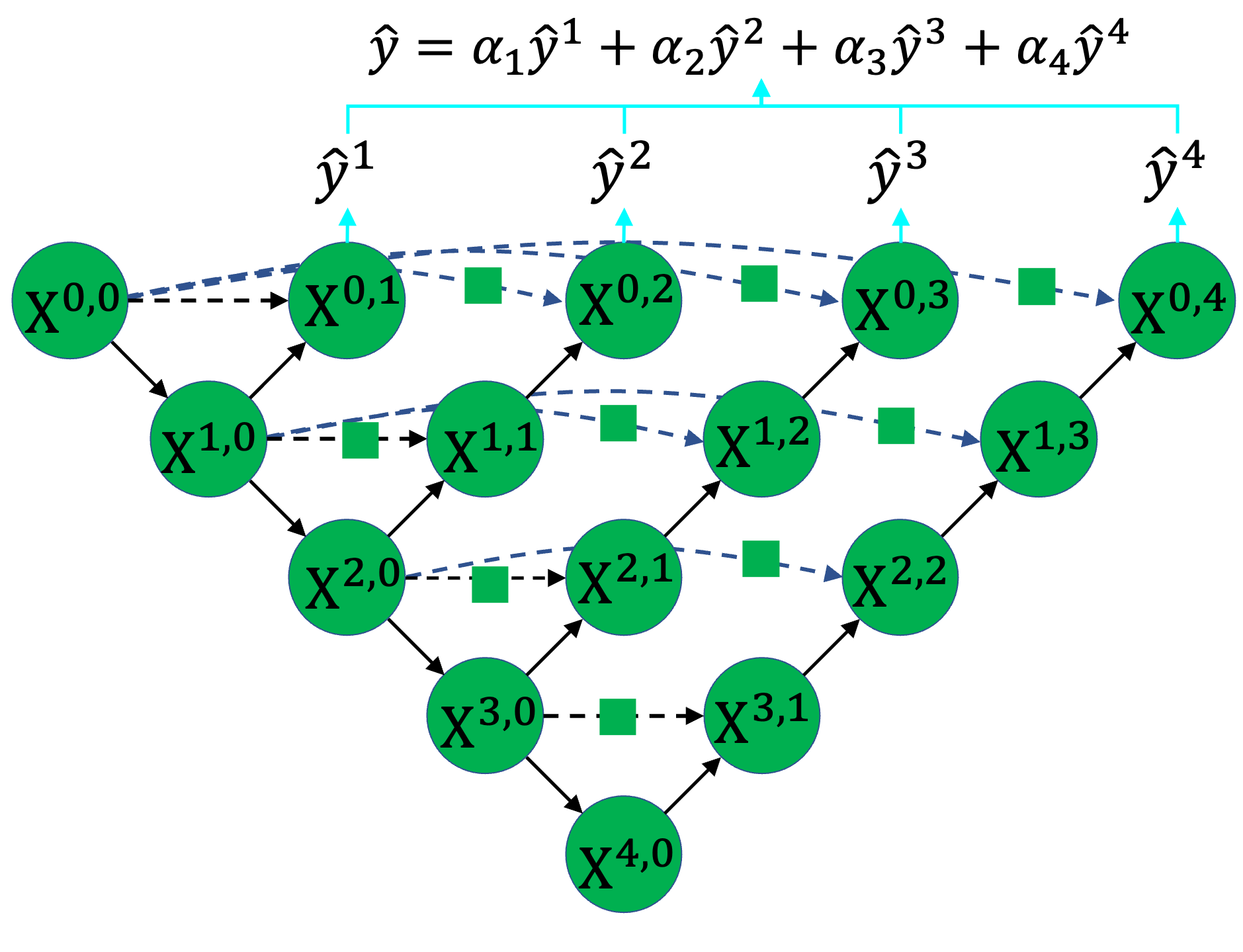}}
	\subfigure[scSE]{
		\label{fig:scSE}
		\includegraphics[scale=0.6]{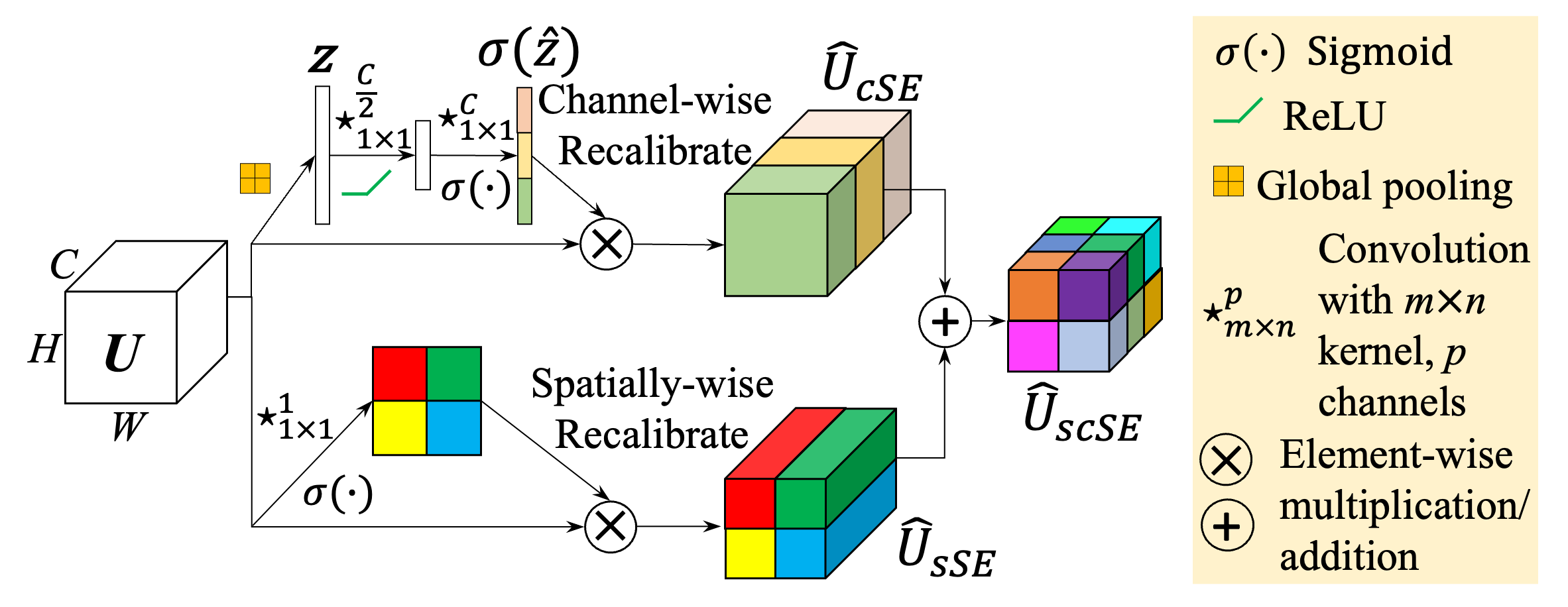}}
	\subfigure[Legend]{
		\label{scSE}
		\includegraphics[scale=0.55]{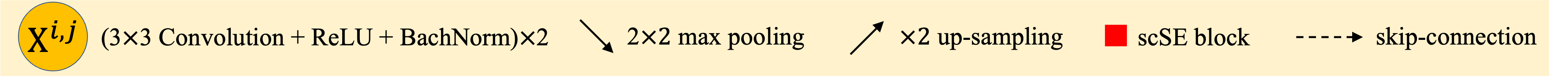}}
	\caption[The architecture of the proposed ADS\_UNet.]{The architecture of the proposed ADS\_UNet. Each circular node in the graph represents a convolution block. Specifically, yellow nodes indicate that parameters are trainable during back-propagation, green nodes indicate that parameters are frozen. (a-d) UNets of varying depths. All of UNet$^i$ are trained with the same dataset $\mathcal X$, but using different sample weight, $W$. (e) Ensemble architecture, ADS\_UNet, which combines UNets of varying depths into one unified architecture for inference. (f) The scSE block. It contains \textit{left branch} channel squeeze and spatial excitation block (sSE), and \textit{right branch} spatial squeeze and channel excitation block (cSE).}
	\label{ITERATION}
\end{figure*}
\subsection{Computation and Memory Efficient Deep Supervision}
\label{section:DS_UNet}
As we mentioned in the introduction section, the UNet$^e$ and UNet++ \citep{zhou2019unet++} offer deep supervision to shallower layers by gradually up-sampling feature maps to the size of the mask, which is computation and GPU memory expensive. To reduce the computational burden, we average-pool the mask to have the same size as feature maps. The advantage of this change is that we no longer need to train deconvolutional weights for intermediate blocks to obtain feature maps with the same dimension as the ground-truth mask. This is of potential benefit for texture segmentation, as relevant textural characteristics occur at multiple length scales, and is not confined to the location of the mask boundary. We adopted UNet$^d$s, whose hidden layers have been trained with supervision, as base learners of the proposed ensemble model. Given the input image $x$ and the network, we define the probability map generated at block X$^{i,j}$ as:
\begin{equation}
    \label{equ:block_y_hat}
    \hat{y}^{i,j}(x)={\rm softmax}({\rm X}^{i,j}(x))
\end{equation}
The mapping X$^{i,j}(\cdot): \bm X\rightarrow \mathbb{R}^{N^{i,j}\times C}$ consists of a sequence of convolution, batch normalization, ReLu activation and pooling operations, to transform the input image to a feature representation. Then a \textit{softmax} activation function is used to map the representation to a probability map. Here $C$ is the number of classes, $N^{i,j}$ denotes the number of pixels of the down-sampled mask, $(i,j)$ denotes the index of convolutional blocks. Given mask $y^{i,j} \in \mathbb{R}^{N^{i,j},C}$, the loss function used in the block X$^{i,j}$ is the pixel-wise cross-entropy loss, which is defined as:
\begin{equation}
    \label{equ:ce-loss}
    \mathcal{L}^{i,j}(y^{i,j},\hat{y}^{i,j},N^{i,j})=-\frac{1}{N^{i,j}}\sum_{n=1}^{N^{i,j}}\sum_{c=1}^{C}{y_{n,c}^{i,j}{\rm log}(\hat{y}_{n,c}^{i,j})},
\end{equation}
where $y^{i,j}_{n,c}$ is the ground-truth label of a pixel and $\hat{y}_{n,c}^{i,j}$ is the probability of the pixel being classified as class $c$. Based on equation~\ref{equ:ce-loss}, the overall loss function for the deep supervised UNet$^d$ is then defined as the weighted sum of the cross entropy loss from each supervised block X$^{i,j}$:
\begin{equation}\label{equ:overall-ce-loss}
    \mathcal{L}_d=\sum_{i,j\geq 0}^{i+j=d}{\eta_{i,j}^d\mathcal{L}^{i,j}(y^{i,j},\hat{y}^{i,j},N^{i,j})},\;\eta^d_{i,j}\geq 0,\;\sum_{i,j\geq 0}^{i+j=d}\eta^d_{i,j}=1,
\end{equation}
where $\eta^d_{i,j}$ is a weighting factor assigned to the convolutional block X$^{i,j}$ to characterize the relative importance of blocks. $d$ denotes the depth of the UNet. In contrast to previous works \citep{dou20163d, zhu2017deeply, zhou2019unet++} that use equal weights $\eta^d_{i,j}=\frac{1}{d+1}$, we initialize $\eta^d_{i,j}$ to $\frac{1}{d+1}$ and allow the $\eta^d_{i,j}$ to be trainable, and use the \textit{softmax} function to normalize ${\eta^d_{i,j}}$ to guarantee $\sum \eta^d_{i,j}=1$. However, the feature learning of a block will be restricted if its $\eta^d_{i,j}$ decreases to 0, during training. In order to guard against this competition exclusion phenomenon and encourage all supervised blocks to contribute to the segmentation, we add a constant $\frac{1}{d+1}$ to $\eta^d_{i,j}$ to raise its lower limit:
\begin{equation}\label{equation:constraint}
    \tilde{\eta}^d_{i,j} = \frac{\eta^d_{i,j}+\frac{1}{d+1}}{\sum_{i,j\geq 0}^{i+j=d}{(\eta^d_{i,j}+\frac{1}{d+1})}} = \frac{\eta^d_{i,j}}{2}+\frac{1}{2(d+1)},
\end{equation}
Since $\lim\limits_{\eta^d_{i,j} \to 0}{\frac{\eta^d_{i,j}}{2}+\frac{1}{2(d+1)}}=\frac{1}{2(d+1)}$ and $\lim\limits_{\eta^d_{i,j} \to 1}{\frac{\eta^d_{i,j}}{2}+\frac{1}{2(d+1)}}=\frac{d+2}{2(d+1)}$, $\tilde{\eta}^d_{i,j}$ is bounded in $[\frac{1}{2(d+1)}, \frac{d+2}{2(d+1)}]$. Then equation~\eqref{equ:overall-ce-loss} is re-written as follows to train each constitute model UNet$^d$:
\begin{equation}\label{equ:overall-ce-loss_2}
\begin{aligned}
    \mathcal{L}_d=\sum_{i,j\geq 0}^{i+j=d}{\tilde{\eta}_{i,j}^d\mathcal{L}^{i,j}(y^{i,j},\hat{y}^{i,j},N^{i,j})},\\
    \frac{1}{2(d+1)} \leq \tilde{\eta}_{i,j}^d \leq \frac{d+2}{2(d+1)}, \sum_{i,j\geq0}^{i+j=d}\tilde{\eta}_{i,j}^d=1.
\end{aligned}
\end{equation}
Once the UNet$^d$ is trained, the final probability map generated by UNet$^d$ is calculated by:
\begin{equation}
    \label{equ:unetd_y_hat}
    \hat{y}^d(x)=\sum_{i,j\geq 0}^{i+j=d}{\tilde\eta^d_{i,j}\hat{y}^{i,j}}(x),
\end{equation}
with $\hat{y}^{i,j}(x)$ and $\eta^d_{i,j}$ defined in equations~\eqref{equ:block_y_hat} and ~\eqref{equation:constraint}. $\hat{y}^d(x)$ denotes the combined prediction of model UNet$^d$. We conduct ablation studies in section \ref{section:analysis_1} to show the benefits of imposing range constraint on $\tilde{\eta}_{i,j}^d$. Moreover, we demonstrate that generating the final prediction by using the weighted summation of multi-scale outputs yields better segmentation performance.
\subsection{Stage-wise Additive Training}\label{sec:iteration}
The stage-wise additive training process of the ADS\_UNet is described in Algorithm~\ref{algorithm:AdaBoost_UNet} and visually illustrated in Figure~\ref{ITERATION}. The main components of the iterative training procedure are 1) updating sample weights, 2) assigning weighting factors to base learners, and 3) freezing trained encoders while training decoders. We will elaborate on these as follows.

Firstly, given the training images $\mathcal X$=$\{x_1,...,x_m\}$ of $n$ pixels each, and associated masks $\mathcal Y$=$\{y_1,...,y_m\}$, we assign a weight $w_k$ to each sample $x_k$. These weights are initialized to $w_k^1=\frac{1}{m}$ (line 1 in Algorithm~\ref{algorithm:AdaBoost_UNet}). Then, in the first iteration ($d$=1), the parameters of the encoder block (X$^{0,0}$) of the first base learner UNet$^1$ are initialized (line 2). In the first iteration of the sequential learning approach, parameters of the bottleneck node X$^{1,0}$ and decoder nodes X$^{0,1}$) of the UNet$^1$ are initialized randomly (lines 4-6). Line 7 initializes the weighting factors $\tilde{\eta}^d_{i,j}$ of supervised blocks.
The UNet$^1$ is then trained on all training samples with the same weight of $\frac{1}{m}$ (line 8). After the UNet$^1$ is trained, the training set will be used to evaluate it and to determine its error rate $\epsilon_1$ (lines 9-11). In contrast to AdaBoost, we use mean Intersection over Union (mIoU) error (lines 10) to measure segmentation performance rather than using mis-classification rate. In detail, given one-hot mask $\bm y_{k,c}$=$[k_1,\cdots,k_n]$, $k_j$$\in$$\{0,1\}$ for a pixel of image $k$ belonging to class $c$ and the corresponding one-hot prediction $\hat{\bm y}_{k,c}^d$=$[\hat{k}_1,\cdots,\hat{k}_n], \hat{k}_j$$\in$$\{0,1\}$ generated by UNet$^d$, the mIoU score $s_k^d$ is calculated by:
\begin{equation}
    \label{equation:mIoU}
    s_k^d = {\rm mIoU}(y_k,\hat{y}_k^d)=\frac{1}{C}\sum_{c=1}^C\frac{\bm y_{k,c}\cdot \hat{\bm y}^d_{k,c}}{\bm y_{k,c}\cdot \bm y_{k,c}+ \hat{\bm y}^d_{k,c}\cdot \hat{\bm y}^d_{k,c}-\bm y_{k,c}\cdot \hat{\bm y}^d_{k,c}},
\end{equation}
where $i$ is the index of training images, $c$ is the index of class labels, $d$ is the index of iteration and also denotes the depth of the constituent UNet. If the error rate $\epsilon_1$ of the UNet$^1$ is less than 1-$\frac{1}{C}$ (line 12), then UNet$^1$ will be preserved for the ensemble, otherwise, it will be disregarded by setting its weighting factor to 0 (lines 18-19). In the case that $\epsilon_1$<$1-\frac{1}{C}$, the equation shown in line 13 is used to calculate model weight $\alpha_d$ for the ensemble. So far we have obtained the first base learner UNet$^1$, and its weighting factor $\alpha_1$.

We then update sample weights based on mIoU scores (line 14) for the training of the next iteration:
\begin{subequations}
    \begin{align}
    \label{equation:w_i_1}
    w_k^d = w_k^{d-1}  e^{(1-s_k^{d-1})},\quad k=1,2,...,m, \\
    \label{equation:w_i_2}
    w_k^d \leftarrow \frac{w_k^d}{\sum_{i=1}^m{w_i^d}},\quad k=1,2,...,m,
    \end{align}
\end{subequations} 
Equation~\eqref{equation:w_i_1} assigns greater weight to images that cannot be accurately segmented by UNet$^{d-1}$, encouraging UNet$^d$ to focus more on their segmentation. Equation~\eqref{equation:w_i_2} normalizes sample weights to guarantee that $\sum_{k=1}^m w_k^d=1$.
\begin{algorithm}[!t] \label{algorithm:AdaBoost_UNet}
\caption{ADS\_UNet. The $\tilde\eta^t_{i,j}$ term in line 9 is discussed in the context of equation~\eqref{equation:constraint}; the UNet$^t$ are described in Figure~\ref{ITERATION}.}
\LinesNumbered 
\KwIn{Number of class: $C$; Training images: $\mathcal X=\{x_1,...,x_m\}$; Training masks: $\mathcal Y=\{y_1,...,y_m\}$; Number of iteration: $T$.}
$\bm w^1 = \{w_k^1 | w_k^1=\frac{1}{m}, k=1, 2,...,m$\} \;
Initialising convolutional block $X^{0,0}$ \;
\For{$d=1,2,...,T$}{
    \For{$j=0,1,...,d$}{
   Initializing convolutional block $X^{d-j,j}$\;
    }
    $\tilde{\eta}^d_{i,j}=\frac{1}{d+1},\; i,j\geq0,\; i+j=d$ \;
    Train UNet$^d(\mathcal X, \mathcal Y, \bm w^d)$ \; 
    $\hat{y}_k^d=\sum_{i,j\geq 0}^{i+j=d}{\tilde{\eta}^d_{i,j}\hat{y}^{i,j}}(x_k)$ \tcp*{see equation~\eqref{equ:unetd_y_hat}}
    $ s_k^d=\mbox{mIoU}(\hat{y}_k^d, y_k)$ \tcp*{see equation~\eqref{equation:mIoU}}
    $\epsilon_d=\sum_{k=1}^{m}{w_k^d (1-s_k^d)}$ \; 
    \eIf{$\epsilon_d < 1-\frac{1}{C}$}{
  $\alpha_d=\frac{1}{2}\ln(\frac{1-\epsilon_d}{\epsilon_d})+\ln(C-1)$ \; 
  Updating sample weight $w_k^d$ using equation~\eqref{equation:w_i_1} and ~\eqref{equation:w_i_2}\;
  \For{$j=0,1,...,d$}{
      Freeze  convolution block $X^{j,0}$ \;
  }
    }{
  $\alpha_d=0$ \;
    }
}
\KwOut{${\rm ADS\_UNet}=\underset{C}{\arg\max}(\sum_{d=1}^{T} \alpha_d \hat{y}^d)$}
\end{algorithm}

Before the start of the second iteration, it is necessary to freeze the encoder nodes (X$^{0,0}$ and X$^{1,0}$) of the UNet$^1$ (lines 15-17). Otherwise, the process of training UNet$^2$ would update UNet$^1$'s encoder parameters as well, reducing the learned association between the encoder and decoder paths of UNet$^1$. Furthermore, subsequent sub-networks $\rm{UNet}^d$, for $d\geq2$ would acquire correlated features. The code block in lines 4-20 is run for $T$ iterations to obtain $T$ base learners, each weighted by $\alpha_d$. Note that all parameters of UNet$^{1}$ are trained as a whole but UNet$^2$ reuses encoder weights of UNet$^{1}$ and only its decoder parameters are trained (if $\epsilon_1 \leq 1-\frac{1}{C}$). These are shown as yellow nodes in Figure~\ref{U2}. The purpose of using the updated sample weights $w_k^2$ to train UNet$^2$ is to force the decoder layers of UNet$^2$ (because the connection between X$^{1,0}$ and X$^{2,0}$ only involves max pooling) to learn features dissimilar to those learned by UNet$^1$. This procedure is repeated for each of the base learners UNet$d$, with the additional help of feature normalization to be described next.
\subsection{Feature Re-calibration}\label{section:scse}
The concurrent spatial and channel Squeeze \& Excitation (scSE) block \citep{roy2018concurrent} is used to re-calibrate feature maps learned from encoder blocks of UNet$^d$, to better adapt to features learned from decoder blocks of deeper UNet$^{d+a}$, $a\geq1$ layers. For example, features learned by the encoder block $X^{0,0}$ and the decoder block $X^{1,0}$ of the UNet$^1$ can cooperate well to perform segmentation since their weights are updated in a coordinated end-to-end back-propagation process. In UNet$^2$, however, features produced by $X^{1,1}$ and $X^{1,0}$ (in the same depth) can be very different, since the gradient flow is truncated between block X$^{1,0}$ and $X^{2,0}$. Therefore, although features produced by $X^{0,0}$ used to cooperate well with that of $X^{1,0}$, it is not guaranteed that it can adapt well to that of $X^{1,1}$. Based on this analysis, the scSE block is used to re-weight features before concatenating. We evaluate the role of feature re-calibration in section~\ref{section:ablation}. The detailed process of scSE is illustrated in Figure~\ref{fig:scSE}.

Given an input feature map $ \bm U \in \mathbb R^{H\times W \times C}$, The channel squeeze operation generates a matrix $\bm q \in \mathbb R^{H\times W}$ with matrix elements $q_{i,j}=W_{sq}\cdot U_{i,j,k}$, $W_{sq}\in\mathbb{R}^C$ maps the vector at each location $(i,j)$ into a scalar. This matrix is then re-scaled by passing it through a sigmoid function $\sigma(\cdot)$, which re-weights the input feature map $\bm U$ spatially,
\begin{equation}
    \hat{U}^{sSE}_{i,j,k}=\sigma(q_{i,j}) U_{i,j,k},
\end{equation}
The global average pooling of the feature map over all pixels produces $\bm z$ with components $z_k$,
\begin{equation}
    z_k = \frac{1}{H\times W}\sum_i^H\sum_j^W U_{i,j,k},\quad k=1,2,\dots,C
\end{equation}
This vector, $\bm z$, is transformed to $\bm{\hat{z}}=\bm{W_1}(\rm{ReLU}(\bm{W_2 z}))$, with $\bm W_1 \in \mathbb R^{C\times \frac{C}{2}}$, $\bm W_2 \in \mathbb R^{\frac{C}{2}\times C}$ being weights of two fully connected layers. The range of the activations of $\hat{\bm z}$ are brought to the interval [0, 1], by passing it through a sigmoid function $\sigma(\cdot)$. The input feature map $\bm U$ is then re-weighted by the re-scaled vector, with its $k^{th}$ channel
\begin{equation}
    \hat{\bm U}^{cSE}_k=\sigma(\hat{\bm z}_k)\bm U_k, \bm U_k \in \mathbb R^{H\times W}, 
\end{equation}
In the channel re-calibrated feature maps $\hat{\bm U}^{cSE}$, the channels that are less important are suppressed and the important ones are emphasized. Finally, after concurrent spatial and channel squeeze and excitation (scSE), a location $(i,j,c)$ of the input feature map $\bm U$ is then given higher activation when it gets high importance from both, channel re-scaling and spatial re-scaling.

\begin{figure}[t]
    \centering
    \subfigure[Info flow of UNet$^1$]{
        \label{fig:info_flow_1}
        \includegraphics[scale=0.53]{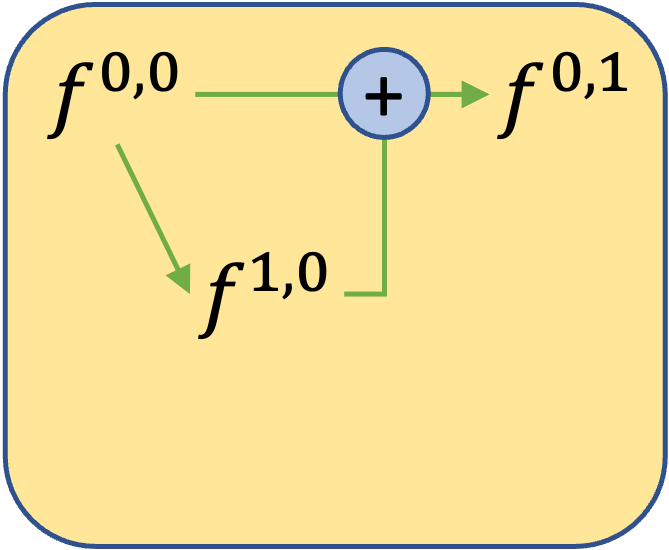}}
    \subfigure[Info flow of UNet$^2$]{
        \label{fig:info_flow_2}
        \includegraphics[scale=0.51]{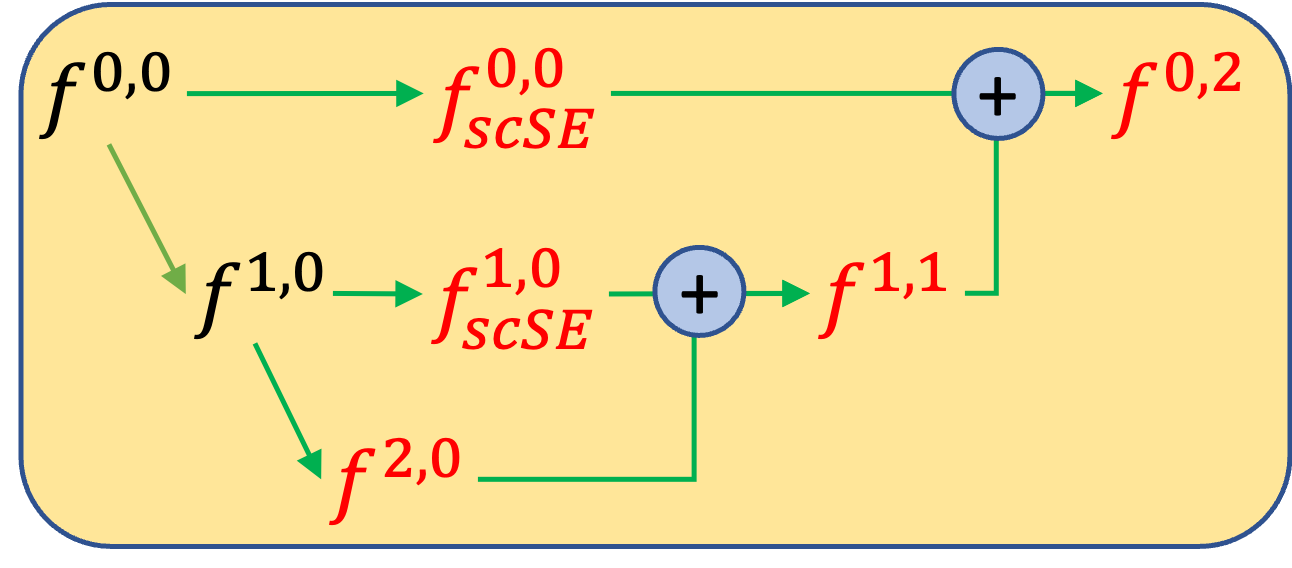}}
    \caption{Information flow diagram of base learners. $f^{i,j}$ denotes the output feature maps of block X$^{i,j}$, $f_{scSE}^{i,j}$ denotes the re-calibrated version of $f^{i,j}$, a circle with plus sign denotes feature map concatenation operation, connecting lines with an arrow denote the flow of features. Showing in red features that can be updated during the training of UNet$^2$ (after UNet$^1$ is trained), while others are fixed.}
    \label{fig:info_flow}
\end{figure}

It is worth mentioning that sample re-weighting and feature re-calibration are utilized in the ADS\_UNet for different purposes and are not in conflict with each other. Taking the pair UNet$^1$ and UNet$^2$ as an example, sample re-weighting aims at achieving feature diversity between final outputs ($f^{0,1}$ and $f^{0,2}$ in Figure~\ref{fig:info_flow}) of two base learners, so that the ensemble of UNet$^1$ and UNet$^2$ can compensate for each other's incorrect predictions, thus leading to better segmentation. When considering feature re-calibration, UNet$^1$ is trained as a whole with each training sample having the same sample weight (as described in section~\ref{sec:iteration}). That means feature maps $f^{0,0}$ and $f^{1,0}$ have a high association. In the second iteration, however, UNet$^2$ reuses UNet$^1$'s encoder blocks (X$^{0,0}$ and X$^{1,0}$ are fixed now), only newly added blocks (X$^{2,0}$, X$^{1,1}$ and X$^{0,2}$) are trained on updated sample weights. This reduces the association of $[f^{1,0}, f^{2,0}]$ and $[f^{0,0}, f^{1,1}]$, enabling feature de-correlation between fixed and newly added feature maps. Directly concatenating $f^{1,0}$ with $f^{2,0}$ and f$^{1,1}$ with $f^{0,0}$ ignores the feature dependence issue and results in worse performance (validated in Table~\ref{tab:ablation_study}). Therefore, to mitigate this feature mismatching effect, we re-calibrate the fixed features ($f^{0,0}\rightarrow f^{0,0}_{scSE}$, $f^{1,0}\rightarrow f^{1,0}_{scSE}$), before concatenation.
\subsection{Difference between ADS\_UNet and UNet++}
In section \ref{section:DS_UNet}-~\ref{section:scse}, we introduced the components and training scheme of the ADS\_UNet. For inference, the final probability map for an image $x\in\mathbb{R}^{C \times H\times W}$ can be generated by weighted average:
\begin{equation}
\begin{aligned}
    \label{equation:test}
    \hat{y}(x)={\rm ADS\_UNet}(x)&=\sum_{d=1}^{T} \alpha_d \hat{y}^d(x)
\end{aligned}
\end{equation}
Here $C$ is the number of classes. $\hat{y}^d(x)$ is the probability map generated by UNet$^d$, as defined in equation~\eqref{equ:unetd_y_hat} and shown in Figure~\ref{U5}.

The proposed ensemble structure differs from the UNet++ in two ways: one differs in the training method, and the other in the way decisions are made and incorporated into learning. 1) \textit{Embedded vs. isolated training}. The UNet++ is trained in an embedded training fashion where the full UNet++ model is trained as a whole, with deep supervision on the last decoder block $X^{0,i}$ of branch $i$. In the ADS\_UNet, however, each UNet$^d$ is trained by isolating features acquired by the deeper encoder and decoder blocks. Moreover, deep supervision is added to each decoder block of each branch by down-scaling the label masks, rather than solely on the last decoder node of each branch. 2) \textit{Average vs. weighted average voting.} In the ensemble mode of the UNet++, the segmentation results from all branches are collected and then averaged to produce the final prediction. UNet++$(x)=\underset{c\in C}{\arg\max}(\frac{1}{T}\sum_{d=1}^{T} {\rm UNet}^d(x))$, with UNet$^d(x)=\hat{y}^{0,d}$. $T$ is the number of branches of the UNet++. However, the ADS\_UNet takes performance-weighted combinations of the component UNets to create the final segmentation map: ADS\_UNet$(x)=\underset{c\in C}{\arg\max}(\sum_{d=1}^{T} \alpha_d \hat{y}^d)$, with $\hat{y}^d=$UNet$^d(x)$ is calculated from equation~\eqref{equ:unetd_y_hat}. $\alpha_d$ reflects the importance of the UNet$^d$ in the ensemble.
\section{Experiments and Results}  

Three histopathology datasets are used to check the effectiveness of the proposed methods.
\subsection{Datasets}
\noindent\textbf{CRAG dataset}. The colorectal adenocarcinoma gland (CRAG) dataset \citep{awan2017glandular} contains a total of 213 Hematoxylin and Eosin images taken from 38 WSIs scanned with an Omnyx VL120 scanner under 20× objective magnification). All images are mostly of size 1512$\times$1516 pixels. The dataset is split into 173 training images and 40 test images. We resize each image to a resolution of 1024$\times$1024 and then crop it into four patches with a resolution of 512$\times$512 for all our experiments.

\noindent \textbf{BCSS dataset}. The Breast Cancer Semantic Segmentation dataset \citep{amgad2019structured} consists of 151 H\&E stained whole-slide images and ground truth masks corresponding to 151 histologically confirmed breast cancer cases. A representative region of interest (ROI) was selected within each slide by the study coordinator, a medical doctor, and approved by a senior pathologist. ROIs were selected to be representative of predominant region classes and textures within each slide. Tissue types of the BCSS dataset consist of 5 classes (\romannumeral1)tumour, (\romannumeral2)stroma, (\romannumeral3)inflammatory infiltration, (\romannumeral4)necrosis and (\romannumeral5)others. We set aside slides from 7 institutions to create our test set and used the remaining images for training. Shift and crop data augmentation, random horizontal and vertical flips were adopted to enrich training samples. Finally, 3154 and 1222 pixel tiles of size 512$\times$512 were cropped for training and testing, respectively. Weighted categorical cross-entropy loss was used to mitigate class imbalance, with the weight associated with each class determined by $W_c=1-\frac{N_c}{N}$, where $N$ is the number of pixels in the training dataset and $N_c$ is the number of pixels belonging to class $c$.

\noindent \textbf{MoNuSeg dataset}. The MoNuSeg dataset \citep{kumar2019multi} is a multi-organ nucleus segmentation dataset. The training set includes 30 images of size 1000$\times$1000 from 4 different organs (lung, prostate, kidney, and breast). The test set contains 14 images with more than 7000 nucleus boundary annotations. A 400$\times$ 400 window slides through the images with a stride of 200 pixels to separate each image into 16 tiles for training and testing.
\subsection{Baselines and Implementation}
\label{section:implement}
Since our work is mainly based on UNet, UNet$^e$, and UNet++, we re-implement these three models, as well as CENet, to compare with our proposed methods. We also compare the proposed ADS\_UNet with two transformer-based UNet variants, HyLT \citep{luo2021hybrid} and MedFormer \citep{gao2022data}, using the implementation provided by the authors. For a fair comparison, the configuration of the outermost convolutional blocks ($X^{i,0},\;i\in\{0,1,2,3\}$ and $X^{i,j},\; i,j\geq0,\;i+j=4$) of all compared methods are exactly the same as in the original UNet (both the number and size of filters). All inner decoder nodes of UNet$^e$, UNet++ and ADS\_UNet are also exactly the same, and all models have the same hyper-parameters. It is noted that scSE block is not used in UNet, UNet$^e$, UNet++ and CENet, but it is used in the skip-connections of ADS\_UNet. The models are implemented in Pytorch \citep{paszke2019pytorch} and trained on one NVIDIA RTX 8000 GPU using the Adam optimizer \citep{kingma2014adam} with weight decay of 10$^{-7}$ and learning rate initialized at 0.001 and then changed according to the 1cycle learning rate policy \citep{smith2019super}. The cross-entropy loss is used to train all compared models, and ADS\_UNet is trained with the linear combination of loss functions using equation~\eqref{equ:overall-ce-loss_2}. On models with a depth of 4, the number of filters at each level are 64, 128, 256, 512, and 1024, on the CRAG and the BCSS dataset. This setting is consistent with the standard UNet \citep{ronneberger2015u}. However, we change the number of filters to 16, 32, 64, 128, 256 for all models, when trained on the MoNuSeg dataset, as our experimental results show that increasing the number of filters leads to inferior performance. The colour normalization method proposed in \cite{vahadane2016structure} is used to remove stain color variation, before training. We also compare our methods with the state-of-the-art nnU-Net \citep{isensee2021nnu}. Note that the nnUNet automatically decides the depth of the architecture based on its characterization of the properties of the datasets. In our experiments, the nnUNet generated for the BCSS dataset and the CRAG dataset is of a depth of 7, while it is 6 for the MoNuSeg dataset. The officially released nnUNet source code is used in our experiments. 
\begin{table*}[!t]
\caption[Segmentation results of UNet and ADS\_UNet.]{The comparison of models in terms of the number of parameters, computational complexity (measured by FLOPs), required GPU memory, training time (seconds) per epoch, and segmentation performance (measured by mIoU). The FLOPs and GPU consumption are computed with 512$\times$512 inputs. The GPU memory consumption is measured by \textit{nvidia-smi} command (batch size=2). In ADS\_UNet, base learners require a different amount of GPU memory, since they vary in depth and the number of parameters (The total number of trainable parameters of the ADS\_UNet is 35.41 million). The mIoU score of the FCN-8 is computed from the confusion matrix provided in the supplementary material of \cite{amgad2019structured}.}
\label{table:exp_res}
\begin{center}
\begin{tabular}{l|cccc|ccc}
\hline
Net & Params(M) & FLOPs(G) & GPU(GB) & Time(s) & CRAG & BCSS & MoNuSeg\\\hline
FCN-8 \citep{amgad2019structured} & -- & -- & -- & -- & -- & 60.55 & -- \\
UNet \citep{ronneberger2015u} & 31.04 & 218.9 & 5.54 & 771 & 86.87 & 59.41  & 80.12  \\
UNet$^e$ \citep{zhou2019unet++} & 34.92 & 445.2 & 9.80 & 1071 & 86.75 & 58.73  & 81.08 \\
UNet++ \citep{zhou2019unet++} & 36.17 & 514.8 & 9.31 & 1303 & 88.04 & 59.85  & 81.29 \\
nnUNet \citep{isensee2021nnu} & 41.27 & 65.6 & 2.92 & 442 & 88.45 & 60.96  & 80.79\\
CENet \citep{zhou2022contextual} & 35.17 & 471.55 & 5.99 & 713 & 86.85 & 59.45  & 81.67 \\
HyLT \citep{luo2021hybrid} & 42.20 & 329.11 & 16.06 & 1500 & 87.70 & 60.45  & 81.69 \\
MedFormer \citep{gao2022data} & 99.54 & 325.76 & 15.48 & 1337 & 87.92 & 60.26  & \textbf{81.84} \\
ADS\_UNet & \makecell[c]{0.41$\rightarrow$1.63$\rightarrow$\\6.65$\rightarrow$26.72} & \makecell[c]{62.61$\rightarrow$114.80$\rightarrow$\\166.93$\rightarrow$219.04} & \makecell[c]{4.00$\rightarrow$ 4.92$\rightarrow$\\5.40$\rightarrow$5.71} & 453 & \textbf{89.04} & \textbf{61.05} & 81.43 \\\hline
\end{tabular}
\end{center}
\end{table*}

\begin{table*}[!t]
    \centering
    \caption{The difference of the mean rank between ADS\_UNet and compared methods. For each tested dataset, the differences in performance are statistically significant if the difference in the mean rank is greater than the critical distance. Values lower than the critical distance are highlighted in bold, indicating there are no significant differences between those models and ADS\_UNet.}
    \label{tab:statical_analysis}
    \begin{tabular}{ccccccccc}
    \hline
        Dataset & Critical Distance & UNet & UNet$^e$ & UNet++ & nnUNet & CENet & HyLT & MedFormer \\ \hline
        CRAG & 0.830 & 1.941 & 2.109 & 1.813 & \textbf{0.316} & 2.025 & 0.831 & 0.972 \\
        BCSS & 0.300 & 0.474 & \textbf{0.082} & 0.336 & 0.426 & \textbf{0.239} & 0.524 & 0.801 \\
        MoNuSeg & 0.700 & 1.662 & 2.387 & 1.978 & 2.32 & 1.622 & \textbf{0.453} & 2.218 \\ \hline
    \end{tabular}
\end{table*}
\subsection{Results}
\begin{figure*}[!b]
    \centering
    \includegraphics[scale=0.37]{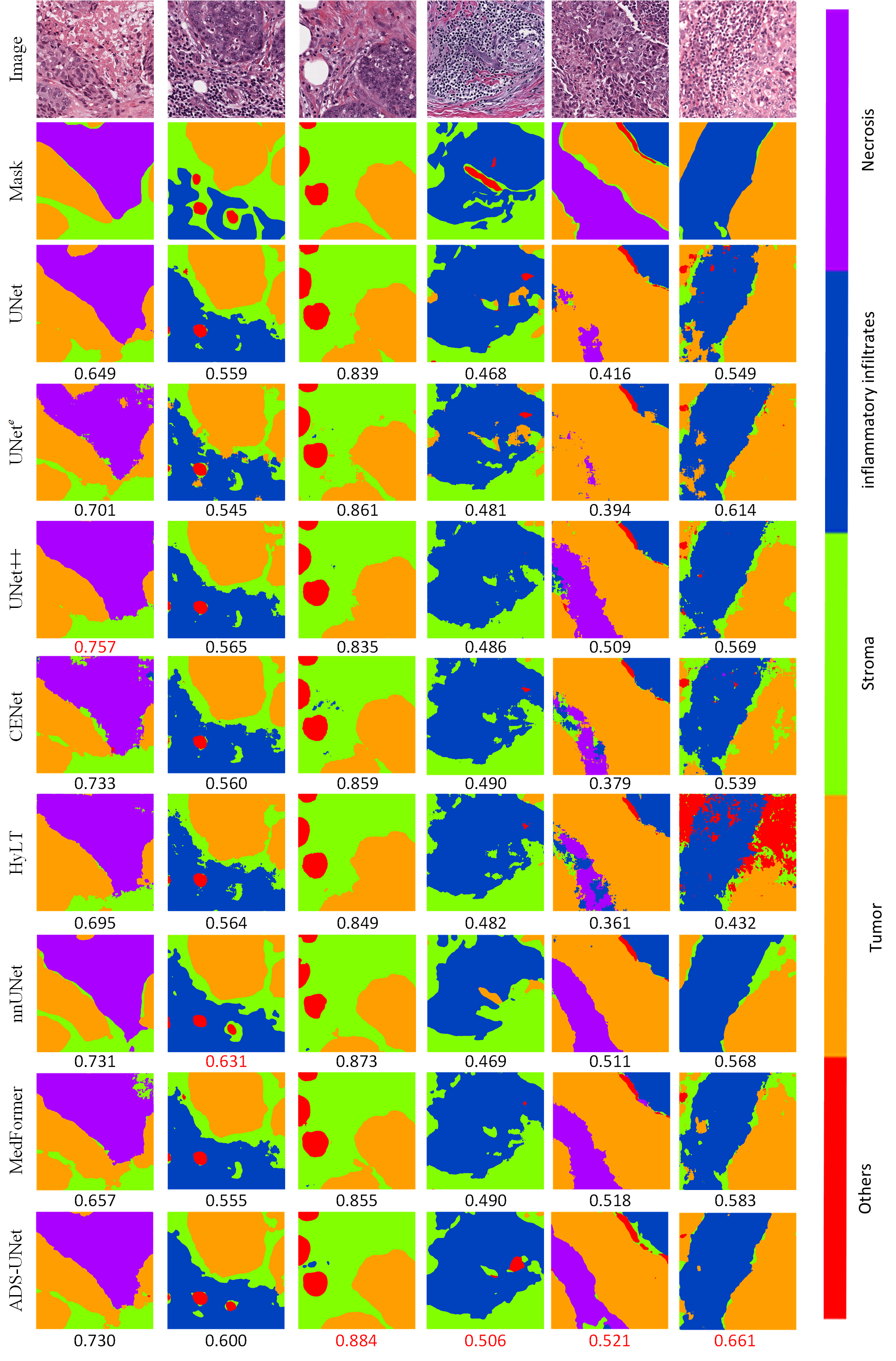}
    \caption[Visual comparison on the BCSS dataset]{Visual comparison of segmentation maps. The mIoU score of each prediction is reported below the prediction.}
    \label{pred_visual}
\end{figure*}

Some image patches and their corresponding segmentation maps are depicted in Figure~\ref{pred_visual}. Table~\ref{table:exp_res} summarizes the segmentation performance achieved by all compared methods. The performance of the baseline method (VGG-16, FCN-8) used in \cite{amgad2019structured} is also included for comparison. The number of parameters and computational efficiency of various UNet variants is also included in the table. Statistical analysis of the results (Table~\ref{table:exp_res}) is performed with the help of the Autorank package \citep{herbold2020autorank}. The non-parametric Friedman test and the \textit{post hoc} Nemenyi test \citep{demvsar2006statistical} at the significance level $\alpha = 0.05$ are applied to determine if there are significant differences between the predictions generated by models and to find out which differences are significant. The performance of ADS\_UNet is compared with 7 other models on 3 datasets. Table~\ref{tab:statical_analysis} shows that 17 of these 21 pairwise comparisons are statistically significant.

\begin{figure}[t]
    \centering
    \subfigure[]{
  \setlength{\fboxsep}{0pt}%
  \fbox{\includegraphics[scale=0.095]{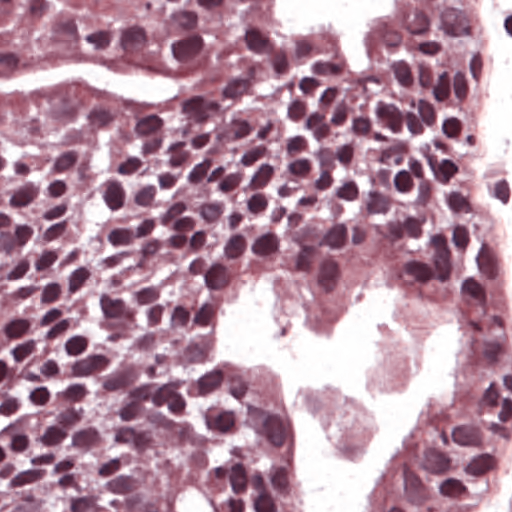}}
    }
    \subfigure[]{
  \setlength{\fboxsep}{0pt}%
  \fbox{\includegraphics[scale=0.095]{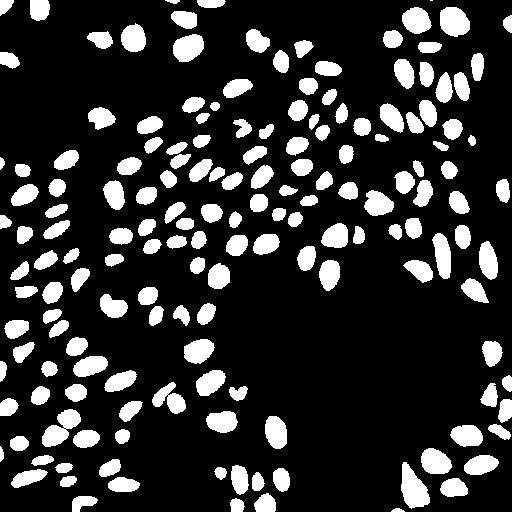}}
    }
    \subfigure[]{
  \setlength{\fboxsep}{0pt}%
  \fbox{\includegraphics[scale=0.095]{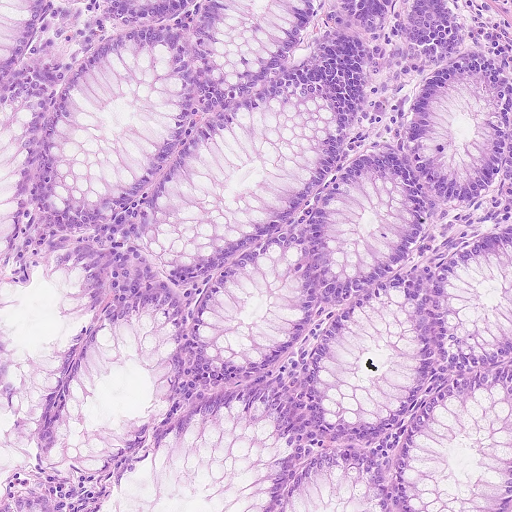}}
    }
    \subfigure[]{
  \setlength{\fboxsep}{0pt}%
  \fbox{\includegraphics[scale=0.095]{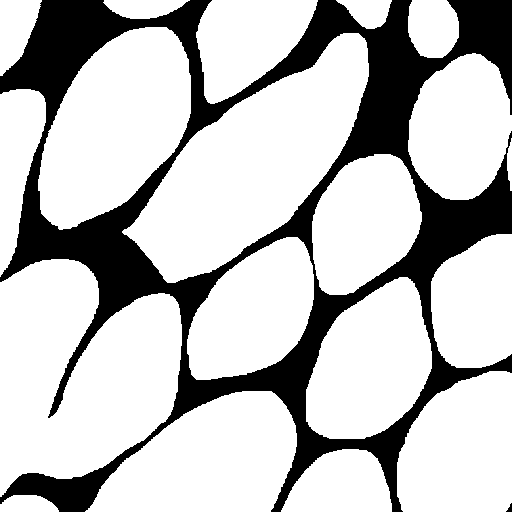}}
    }
    \caption{(a)-(b) Image-mask patch from the MoNuSeg dataset (nucleus segmentation). (c)-(d) Image-mask patch from the CRAG dataset (gland segmentation). All patches are of size $512\times 512$.}
    \label{fig:example-patch}
\end{figure}
Among the different networks evaluated, the ADS\_UNet outperforms all of the other state-of-the-art approaches on the CRAG and BCSS datasets, and achieves competitive performance on the MoNuSeg dataset. UNet++ achieves 1.17, 0.44 and 1.17 higher mIoU scores than UNet by performing 2.35 times more computation and consuming 1.77 times more GPU memory. In contrast, ADS\_UNet performs the best and yet requires at most 59.51\% of the GPU memory and 42.55\% of the floating-point operations required by UNet++ for training. CENet surpasses ADS\_UNet on the MoNuSeg dataset, but at a cost of requiring 2.15 times more computation and 1.19 times more GPU memory. nnUNet consumes the least amount of GPU memory and the number of operations, but at the cost of a small decrease in segmentation accuracy. The design choices (pipeline fingerprint) of nnUNet are not fixed across datasets, but are configured on the fly according to the ‘data fingerprint’ (dataset properties such as image size, image spacing, number of classes, etc.). The data-dependent ‘rule-based parameters’ (patch size, batch size, network depth, etc.) of the pipeline is determined by a set of heuristic rules that models parameter inter-dependencies. As shown in Table~\ref{table:exp_res}, nnUNet outperforms all models on CRAG and BCSS datasets, except for ADS\_UNet. But it demonstrates inferior performance on the MoNuSeg dataset. This can be explained by the characteristics of datasets and the receptive field size of models. Firstly, the nnUNet is deeper (the depth of the nnUNet is 6 or 7, as mentioned in section ~\ref{section:implement}), which means that the convolutional kernels of the bottleneck layer (the deepest encoder layer) have a larger receptive field, enabling the model to extract information from a larger region. This is especially beneficial when the task is to recognize large objects, e.g. tissue types or glands, since a larger receptive field can cover the whole object. In models with a depth of 4, the size of the receptive field of the bottleneck layer is limited. This difference in the depth of models may explain why nnUNet outperforms shallower models when trained for segmenting tissues and glands. In contrast, the size of the cell nucleus in the MoNuSeg dataset is much smaller than tissue and gland. The receptive field of the bottleneck layer of shallow models is large enough to capture the entire nucleus. Further increasing the depth of the network compresses features leading to information loss rather than enhancing the features learnt. The nnUNet improves segmentation performance by enlarging receptive field size, while ADS\_UNet achieves so by ensembling. Image and mask patches presented in Figure~\ref{fig:example-patch} show the size difference of target objects between the nucleus segmentation dataset and the gland segmentation dataset. 

Both transformer-based architectures, HyLT and MedFormer, demonstrate inferior performance on the CRAG dataset, but achieve competitive performance on the BCSS dataset, and outperform the ADS UNet on the MoNuSeg datasets. However, it is worth noting that the HyLT and the MedFormer have 1.19 times and 2.81 times parameters than the ADS\_UNet does and require 2.81 fold and 2.71 fold increases in GPU memory than the ADS\_UNet does for training. The high demand for GPU memory in the HyLT and MedFormer is not surprising, as the attention blocks introduce extra intermediate feature maps that should be kept in the GPU memory for back-propagation.

The amount of computation  (FLOPs) and GPU memory requirement are the main constraints on training speed. Among all compared methods, ADS\_UNet shows a clear advantage in training speed, because the lower GPU memory requirement of ADS\_UNet allows us to use a larger batch size for faster training. The training speed of nnUNet is close to ADS\_UNet, for the same reason. The transformer-based models (MedFormer and HyLT) are the slowest ones since they have the highest GPU memory demand and relatively high computation cost.

\begin{table*}[!t]
    \centering
    \caption{Comparison between the original UNet (without deep supervision) and UNet$^\uparrow$/UNet$^\downarrow$ with deep supervision using up-sampled feature maps/average pooled masks.}
    \begin{tabular}{l|ccc|ccc}
    \hline
  Net & Params (M) & FLOPs (G) & GPU (GB) & CRAG & BCSS & MoNuSeg\\\hline
   UNet & \textbf{31.04} & \textbf{218.9} & \textbf{5.54} & 86.87 & \textbf{59.41} & 80.12 \\
   UNet$^\uparrow$ & 31.17 & 260.27 & 14.03 & \textbf{88.84} & 58.40 & \textbf{81.40} \\
   UNet$^\downarrow$    & 31.06 & 219.08 & 5.61 & 88.33 & \textbf{59.41} & 81.24 \\\hline
    \end{tabular}
    \label{tab:ds_unet_up}
\end{table*}
\begin{table*}[!t]
\caption[Segmentation results of UNet and ADS\_UNet.]{Ablation study. Performance measured by mIoU (highest score highlighted in bold). "SCSE" denotes spatial and channel squeeze \& excitation used in skip-connections. "Re-weight" denote training sample re-weighting. "ens (avg)"/"ens ($\alpha$)" denote that segmentation results from all branches are collected and then averaged, or summed by $\alpha$ weights.} 
\label{tab:ablation_study}
\begin{center}
\begin{tabular}{c|ccc|cccccc}
\hline
Model Name & Deep supervision & SCSE & Re-weight & UNet$^1$ & UNet$^2$ & UNet$^3$ & UNet$^4$ & ens(avg) & ens($\alpha$) \\\hline
model\_0 & \XSolidBrush & \XSolidBrush & \XSolidBrush & 43.41 & 53.66 & 57.59 & 58.40 & 57.92 & 58.22 \\
model\_1 & \checkmark & \XSolidBrush & \XSolidBrush & 43.89 & 53.14 & 57.21 & 58.07 & 58.79 & 58.82 \\
model\_2 & \XSolidBrush & \checkmark & \XSolidBrush & 46.67 & 55.87 & 58.70 & 59.92 & 59.92 & 60.20 \\
model\_3 & \XSolidBrush & \XSolidBrush & \checkmark & 43.58 & 53.93 & 57.34 & 58.43 & 58.59 & 58.82 \\
model\_4 & \XSolidBrush & \checkmark & \checkmark   & \textbf{47.97} & 56.47 & 59.51 & 59.55 & 60.51 & 60.76\\
model\_5 & \checkmark  &  \XSolidBrush & \checkmark & 44.04 & 53.32 & 55.87 & 57.99 & 58.23 & 58.25 \\
model\_6 & \checkmark  &  \checkmark & \XSolidBrush & 46.93 & 56.60 & 58.62 & 60.26 & 60.57 & 60.63 \\
ADS\_UNet & \checkmark  &  \checkmark & \checkmark & 46.93 & \textbf{56.91} & \textbf{60.11} & \textbf{60.26} & \textbf{61.04} & \textbf{61.05} \\\hline
\end{tabular}
\end{center}
\end{table*}
\subsection{Ablation Studies}\label{section:ablation}
\subsubsection{Down-sampling masks vs. up-sampling feature maps}
We build UNet$^\uparrow$ and UNet$^\downarrow$ as UNet’s counterparts to demonstrate the advantage of using down-sampled masks for deep supervision. In UNet$^\uparrow$, feature maps of the UNet are bilinearly interpolated to fit the size of the original mask, while in UNet$^\downarrow$, the masks are down-sampled to fit the size of feature maps. As shown in Table~\ref{tab:ds_unet_up}, UNet$^\downarrow$ with average pooled masks outperforms UNet by 1.46 and 1.12 mIoU on the CRAG and MoNuSeg datasets. This is achieved with only 0.06\% more parameters, 1.26\% more GPU memory consumption and 0.08\% more FLOPs. This small increase comes from a 1$\times$1 convolution layer appended to supervised blocks. We attribute this performance gain to back-propagation through deep layers enforcing shallow layers to learn discriminative features. UNet$^\uparrow$ yields 0.51 and 0.16 higher mIoU than UNet$^\downarrow$ on CRAG and MoNuSeg dataset, but performs worse than UNet by 1.01 points on BCSS dataset. The 18.80\% more computation of the UNet$^\uparrow$ (compared with UNet$^\downarrow$) originates from bilinear interpolation operations when up-sampling feature maps. The GPU memory required in the training process of UNet$^\uparrow$ is 2.50 times that of UNet$^\downarrow$. The reason is that during back-propagation the output of all layers is cached during forward propagation, and the size of the feature map of the supervision layer in UNet$^\uparrow$ is 4 to 256 times the size of the corresponding one in UNet$^\downarrow$. Therefore, beyond a small performance improvement, UNet$^\downarrow$ saves more than 1.50$\times$ GPU consumption thus enabling us to use a larger batch size and save training time.\\
\subsubsection{Tracing the origin of the performance gain of ADS\_UNet.} 
To gain insight into the reason why ADS\_UNet demonstrates superior performance on segmentation, we construct eight models and evaluate them on the BCSS dataset, with each of them being a combination of deep supervision, SCSE feature re-calibration blocks and sample re-weighting. The configuration of models, the performance of each constituent UNet$^d$ and their ensemble performance is summarized in Table~\ref{tab:ablation_study}. To see whether weighted average voting of base learners is better than simple average voting or not, we also compare these two ensemble strategies and report results in Table~\ref{tab:ablation_study}.

\begin{figure*}[!t]
	\centering
  \subfigure[$\Tilde\eta_{i,j}$ values (MoNuSeg).]{
	    \label{fig:eta_DS_monuseg}
    	\includegraphics[scale=0.8]{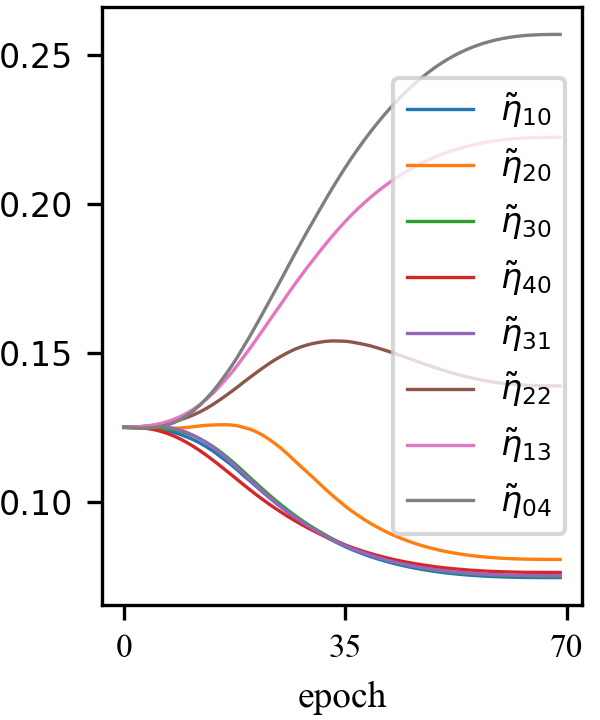}}
  \subfigure[$\Tilde\eta_{i,j}$ values (CRAG).]{
	    \label{fig:eta_DS_crag}
    	\includegraphics[scale=0.8]{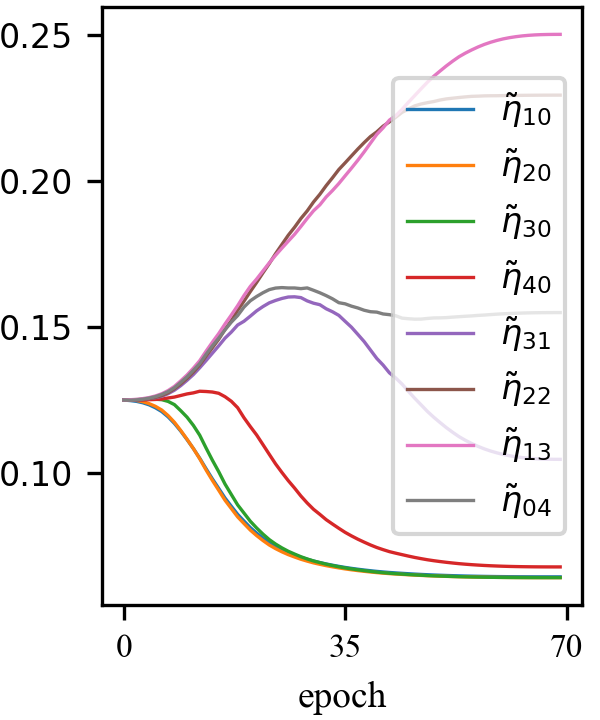}}
  \subfigure[$\Tilde\eta_{i,j}$ values (BCSS).]{
	    \label{fig:eta_DS_bcss}
    	\includegraphics[scale=0.8]{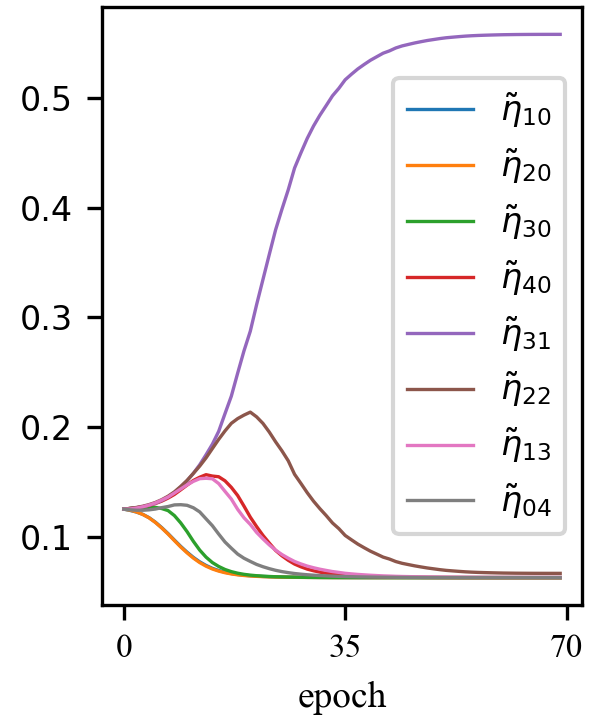}}
  \subfigure[Training losses (BCSS).]{
		\includegraphics[scale=0.8]{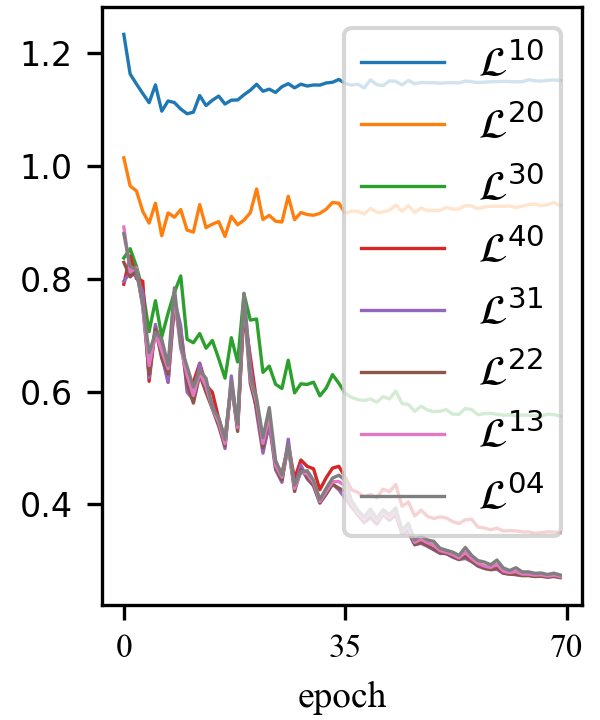}}
    \caption[The loss and the $\bm\eta$ of the UNet's supervision layers.]{(a)-(c) figures show how $\eta^d_{i,j}$ changes when the UNet$^\downarrow$ is trained on the MoNuSeg, CRAG and BCSS datasets. The changing of $\eta_{i,j}$ varies from dataset to dataset. (e) The training losses of supervision layers (trained on the BCSS dataset).}
\end{figure*}
As seen in Table~\ref{tab:ablation_study}, when compared with model\_0 (none of three components is used), model\_1, model\_2 and model\_3 demonstrate the effectiveness of incorporating deep supervision, SCSE feature re-calibration and sample re-weighting into the training of each constituent UNet$^d$, respectively. Moreover, all constitute UNet$^d$s of model\_2 (with SCSE) surpass the counterparts of model\_0, model\_1 and model\_3. This supports the claim we made in section~\ref{section:scse}, namely, features from the encoder block of UNet$^{d-1}$ should be re-calibrated before concatenating with features from the decoder block of UNet$^d$.

Removing deep supervision from the ADS\_UNet drops the mIoU score by 0.29 points (compared with model\_4). Further analysis is provided in section~\ref{section:analysis_1} to reveal the reason why introducing explicit deep supervision leads to better performance.

In model\_4, model\_5 and model\_6, we either remove deep supervision or the SCSE block or sample re-weighting from the ADS\_UNet, respectively, to show the importance of each component in the composition of the ADS\_UNet. As seen in Table~\ref{tab:ablation_study}, removing any one of them would lead to lower segmentation performance.

The experiment conducted on model\_5 demonstrates that truncating the gradient flow between encoder blocks of UNet$^d$ and decoder blocks of UNet$^{d+1}$ is detrimental to the final segmentation performance (compared with the ADS\_UNet). By introducing feature re-calibration in skip-connections, features learnt in encoder blocks are re-weighted to adapt to the ones of decoder blocks, thereby leading to better performance. The importance of SCSE feature re-calibration is also reflected in comparisons of model\_0 vs. model\_2 (1.98$\uparrow$), model\_1 vs. model\_6 (1.81$\uparrow$), and models\_3 vs. model\_4 (1.94$\uparrow$).

In terms of sample re-weighting, the ensemble (ens($\alpha$)) of ADS\_UNet surpasses the one of model\_6 by 0.42 points. We attribute this to sample weight updating, which allows UNet$^d$ to pay more attention to images which are hard to be segmented by UNet$^{d-1}$. The benefit of sample re-weighting is also reflected in comparisons of model\_0 vs. model\_3 (0.6$\uparrow$) and model\_2 vs. model\_4 (0.56$\uparrow$).

When comparing ensemble strategies, we find both average voting and weighted voting improve segmentation performance compared with UNet$^4$; but the improvement due to weighted voting is higher than from average voting. Moreover, the segmentation performance of the model\_6 with $\alpha$ weighting is better than that of average weighting, although training samples are not re-weighted in its iterative training process. This, too, supports the view that integrating multiple models by weighting each as per its segmenting ability improves the overall performance of the ensemble.
\section{Analysis}
\subsection{Incorrect labeling information can be evaded by adjusting $\Tilde{\eta}^d_{i,j}$.}
\begin{table}[!t]
    \centering
    \caption{The proportion of incorrect labels in different scaled masks. The (X$^{i,j}$) under the down-scale factor indicates which layers the mask down-sampled by this down-scale factor is used to supervise.}
    \begin{tabular}{l|cccc}
  \hline
  \diagbox{Data}{$\downarrow$} & \makecell[c]{2\\X$^{1,0}$, X$^{1,3}$} & \makecell[c]{4\\X$^{2,0}$, X$^{2,2}$} & \makecell[c]{8\\X$^{3,0}$, X$^{3,1}$} & \makecell[c]{16\\X$^{4,0}$}  \\\hline
   CRAG    & 1.02 & 2.78 & 5.72 & 9.94 \\ 
   BCSS    & 1.51 & 4.32 & 10.04 & 19.75 \\
   MoNuSeg & 5.49 & 16.21 & 36.69 & 60.05 \\\hline
    \end{tabular}
    \label{tab:incorret_ratio}
\end{table}
It is true that down-sampling the ground-truth mask eliminates small objects and leads to incorrect labels for pixels located on the class boundaries. We quantify the ratio of incorrect labels of down-sampled masks and present the statistics in Table~\ref{tab:incorret_ratio}. It can be observed that the proportion of incorrect labels rises as the down-scaling factor becomes larger. Incorrect labels in the $\times$16 down-scaled mask in the CRAG and BCSS datasets account for 9.94\% and 19.75\% of the total labels, respectively. This figure soars up to 60.05\% in the MoNuSeg dataset. However, it is noteworthy that when these reduced masks are used to supervise the training of layers, there is a trainable weight $\Tilde{\eta}^d_{i,j}$ (defined in equation~\eqref{equation:constraint}) that dynamically adjusts the strength of each layer being supervised. Figure~\ref{fig:eta_DS_monuseg}-\ref{fig:eta_DS_bcss} shows how the network adjusts $\Tilde{\eta}_{i,j}$ during training to assign weightings to layers and scales that contribute most to the segmentation task. As seen, at the end of the training, the largest $\Tilde{\eta}_{i,j}$ values of the MoNuSeg, CRAG and BCSS datasets come from $\Tilde{\eta}_{0,4}$, $\Tilde{\eta}_{1,3}$ and $\Tilde{\eta}_{3,1}$, respectively. That means the UNet$^\downarrow$ benefits most from the original mask and the mask down-scaled by a factor of 2, 8, when trained on the MoNuSeg, CRAG and BCSS dataset. The $\times$2 and $\times$8 down-scaled masks carry 1.02\% and 10.04\% incorrect label information, respectively. Therefore, even though the down-scaled masks introduce wrong labelling information, the UNet$^\downarrow$ is able to evade this wrong information to a certain extent and puts attention on the informative mask by adjusting $\Tilde{\eta}_{i,j}$. Despite the (apparently significant) labelling errors introduced by down-sampling, the overall result (as shown in Table~\ref{tab:ds_unet_up}) is not adversely affected.
\begin{figure*}[!h]
\centering
\subfigure[$\mathcal{L}^{i,j},\;i+j=1$]{
    \label{fig:loss_Ada_const_1}    
    \includegraphics[scale=0.54]{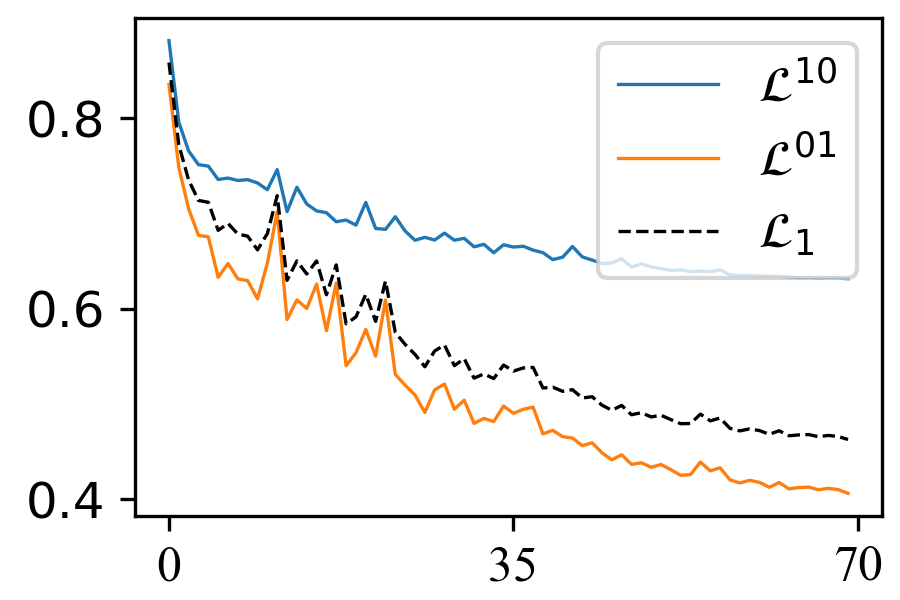}}
\subfigure[$\mathcal{L}^{i,j},\;i+j=2$]{
    \label{fig:loss_Ada_const_2}	    
    \includegraphics[scale=0.54]{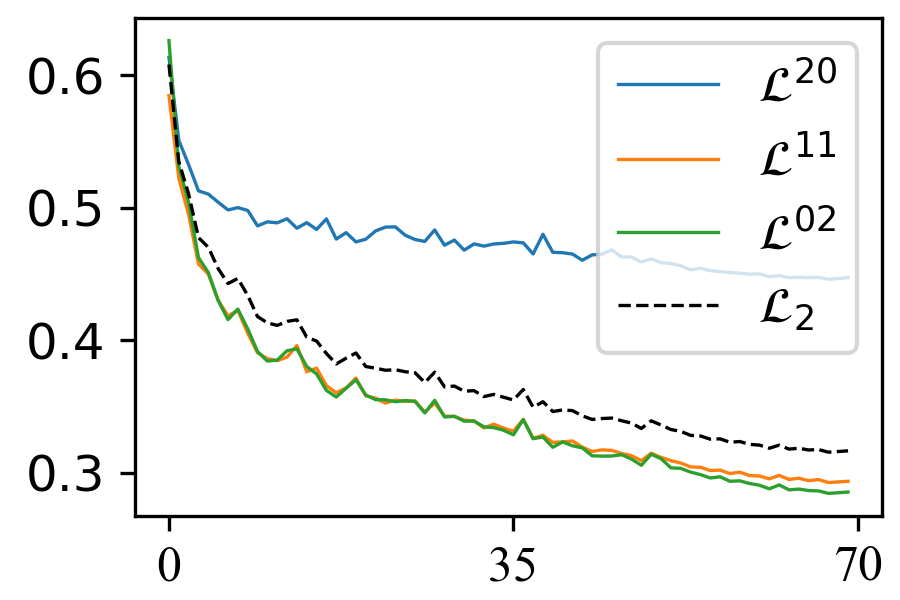}}
\subfigure[$\mathcal{L}^{i,j},\;i+j=3$]{
    \label{fig:loss_Ada_const_3}
    \includegraphics[scale=0.54]{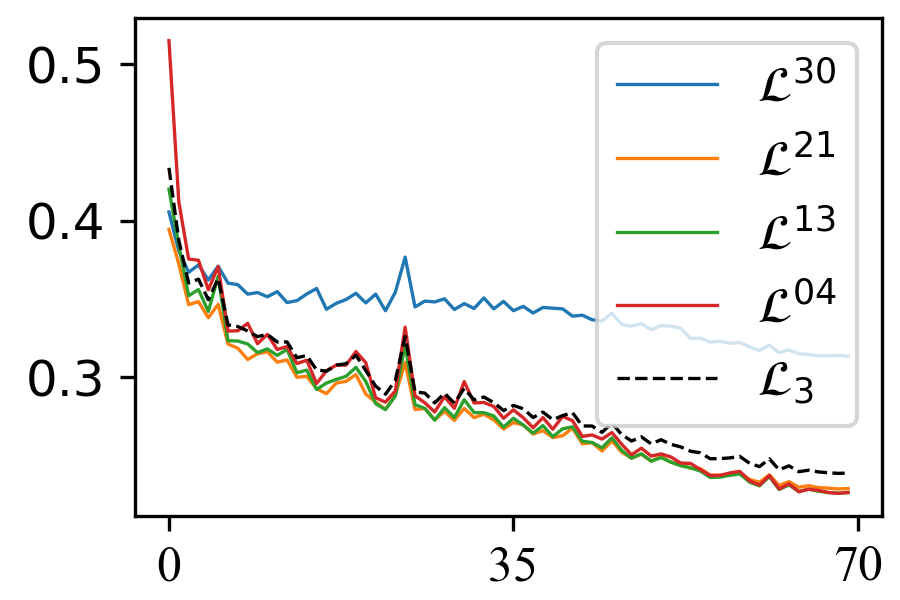}}
\subfigure[$\mathcal{L}^{i,j},\;i+j=4$]{
    \label{fig:loss_Ada_const_4}
    \includegraphics[scale=0.54]{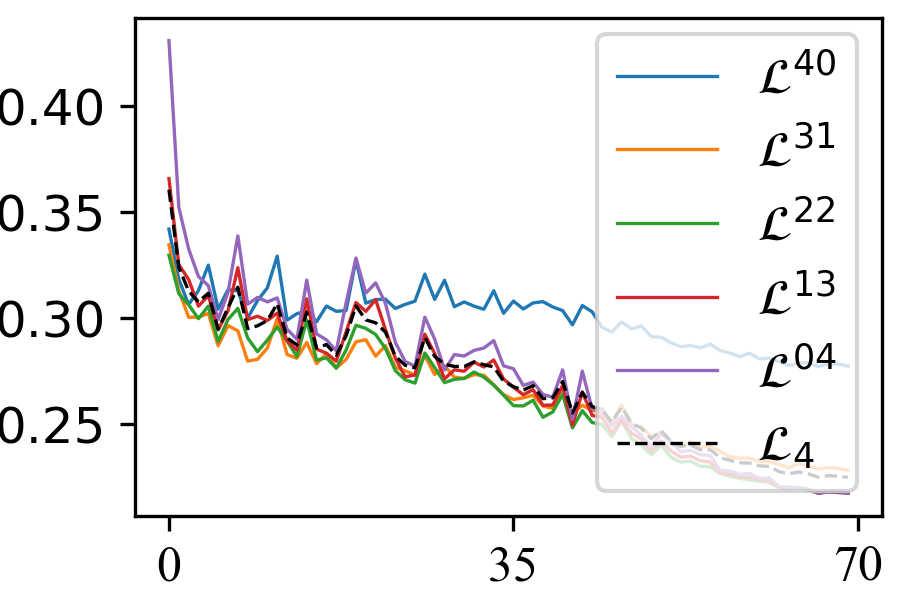}}
\subfigure[$\Tilde\eta^1_{i,j},\;i+j=1$]{
    \label{fig:eta_Ada_const_1}
    \includegraphics[scale=0.54]{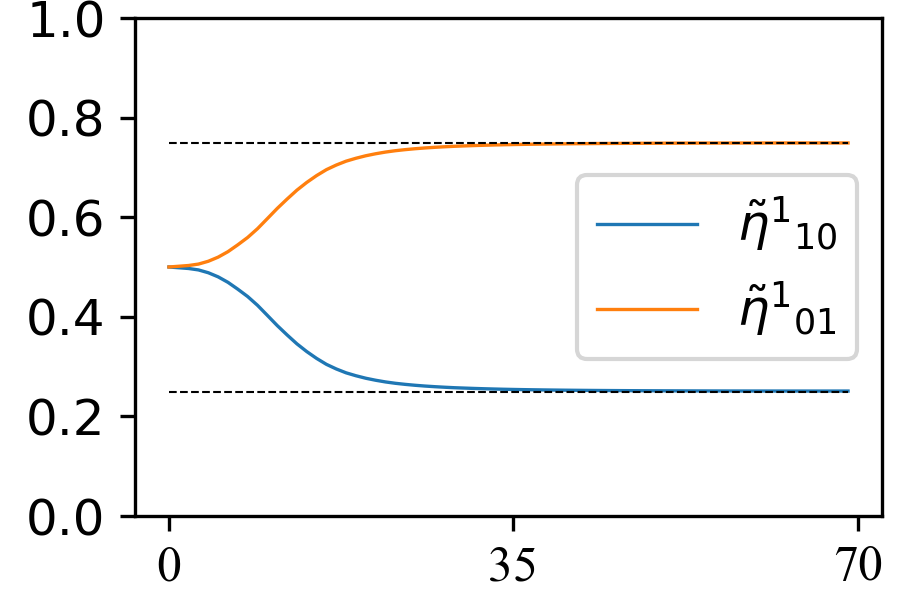}}
\subfigure[$\Tilde\eta^2_{i,j},\;i+j=2$]{
    \label{fig:eta_Ada_const_2}
  \includegraphics[scale=0.54]{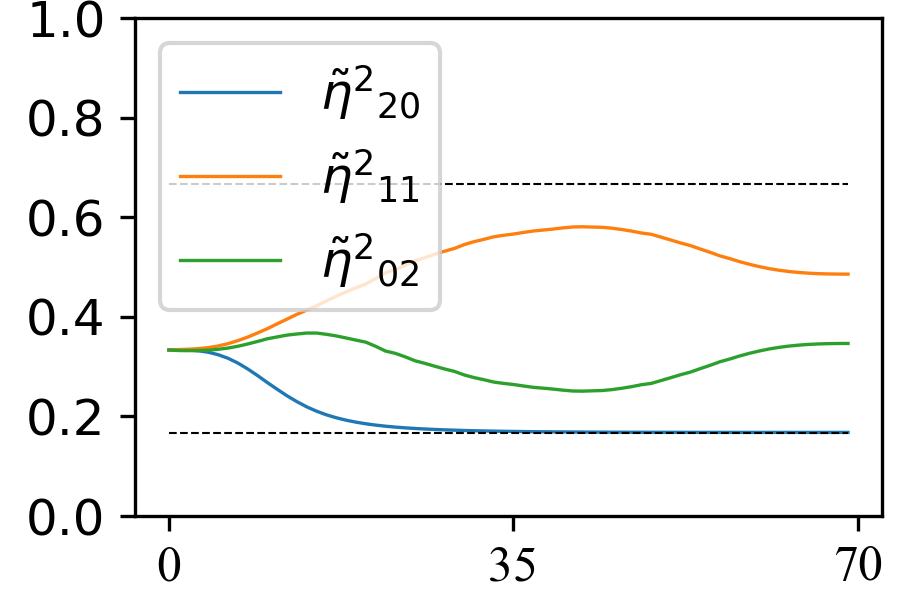}}
\subfigure[$\Tilde\eta^3_{i,j},\;i+j=3$]{
  \label{fig:eta_Ada_const_3}
    \includegraphics[scale=0.54]{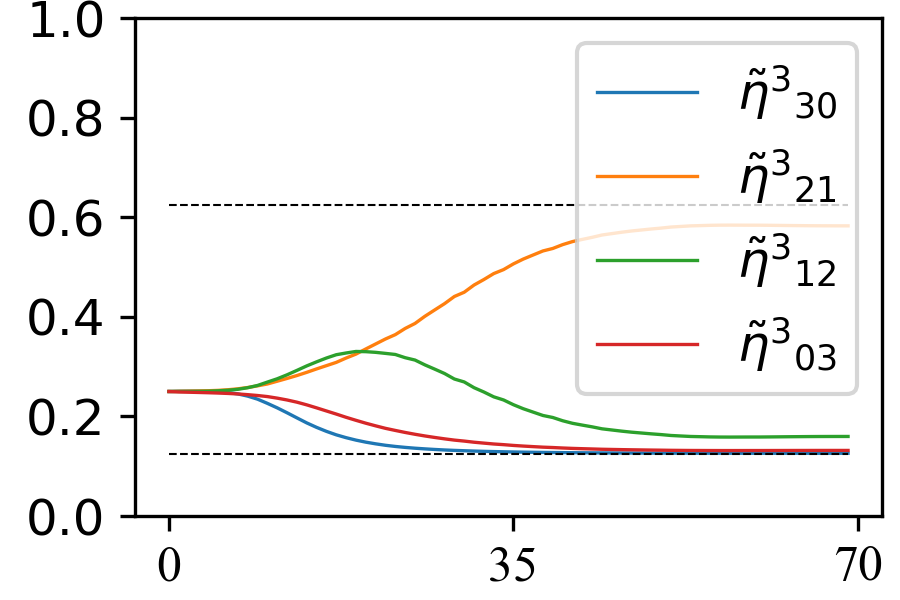}}
\subfigure[$\Tilde\eta^4_{i,j},\;i+j=4$]{
    \label{fig:eta_Ada_const_4}
    \includegraphics[scale=0.54]{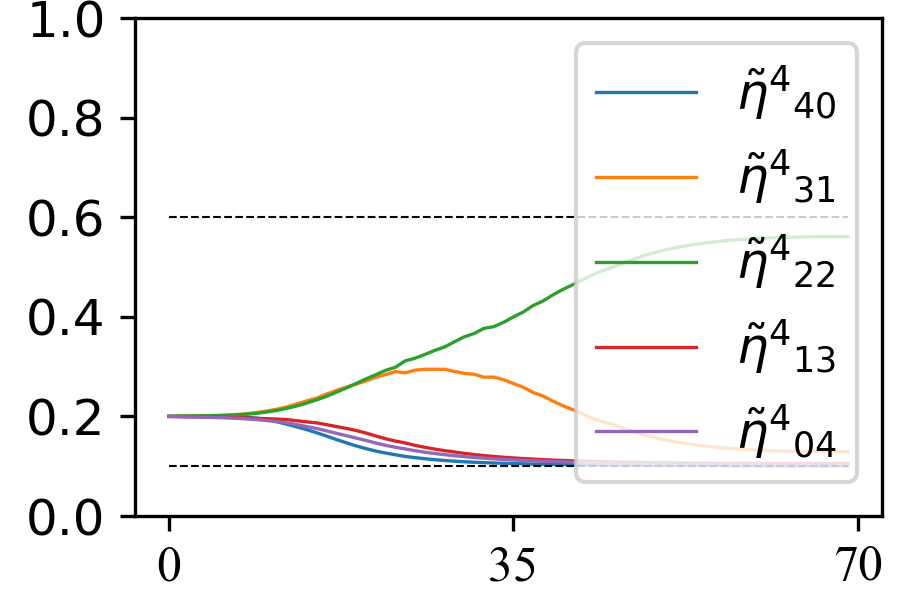}}
\subfigure[$\mathcal{L}^{i,j},\;i+j=1$]{
\label{fig:loss_Ada_1}
    \includegraphics[scale=0.54]{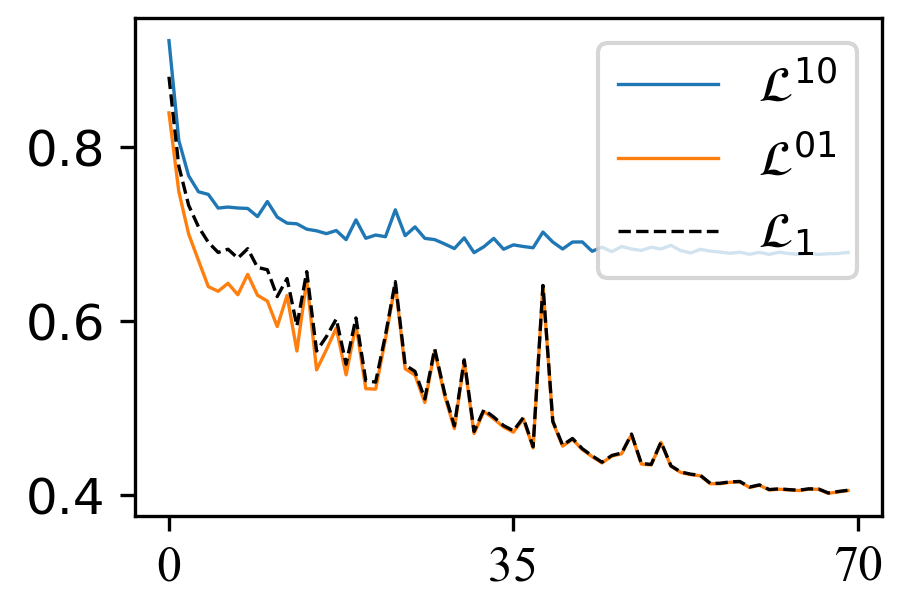}}
\subfigure[$\mathcal{L}^{i,j},\;i+j=2$]{
\label{fig:loss_Ada_2}
    \includegraphics[scale=0.54]{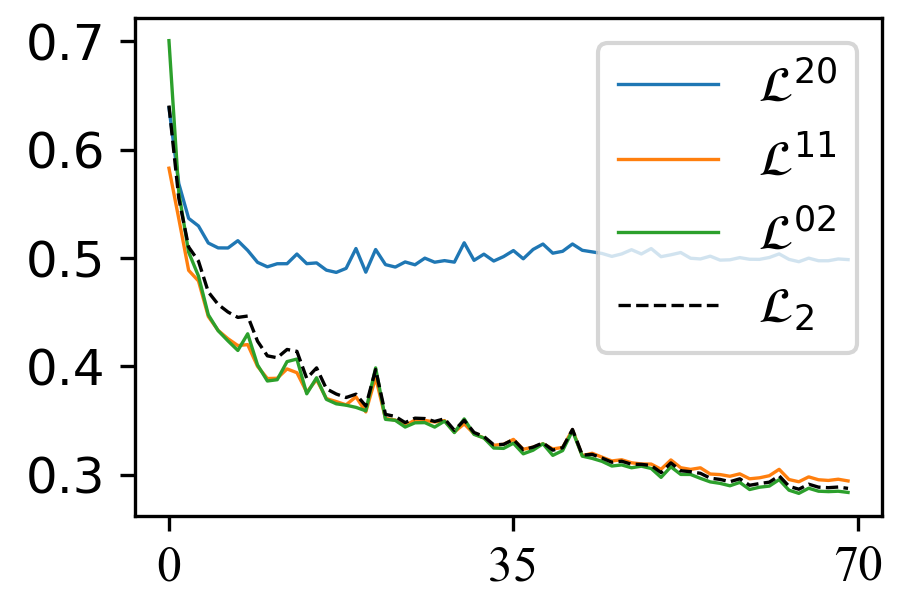}}
\subfigure[$\mathcal{L}^{i,j},\;i+j=3$]{
\label{fig:loss_Ada_3}
    \includegraphics[scale=0.54]{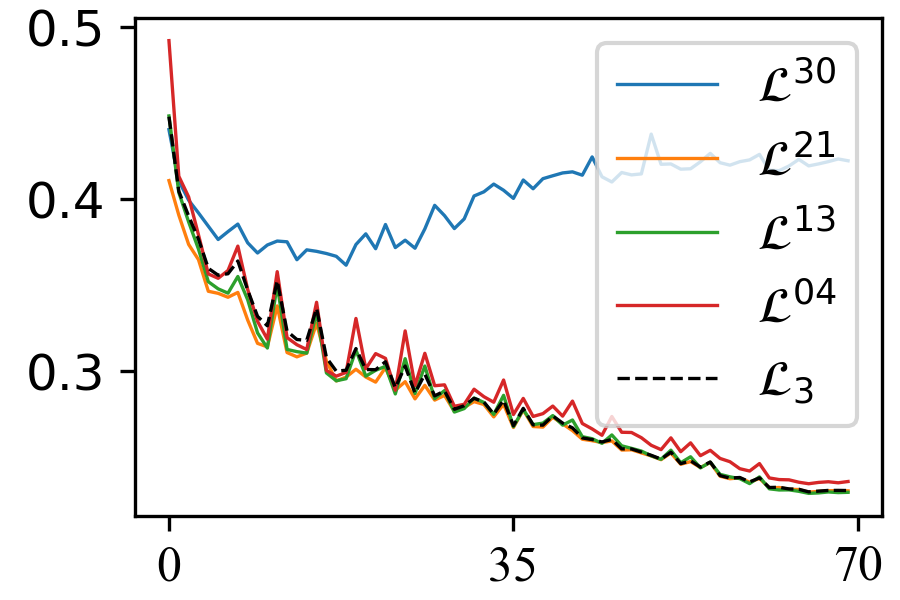}}
\subfigure[$\mathcal{L}^{i,j},\;i+j=4$]{
\label{fig:loss_Ada_4}
    \includegraphics[scale=0.54]{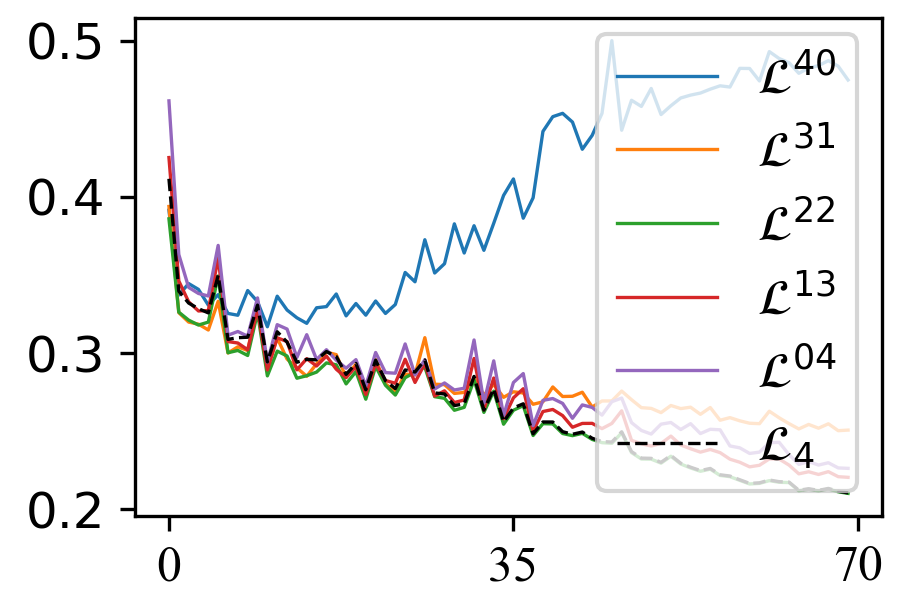}}
\subfigure[$\eta^1_{i,j},\;i+j=1$]{
    \label{fig:eta_Ada_1}
    \includegraphics[scale=0.54]{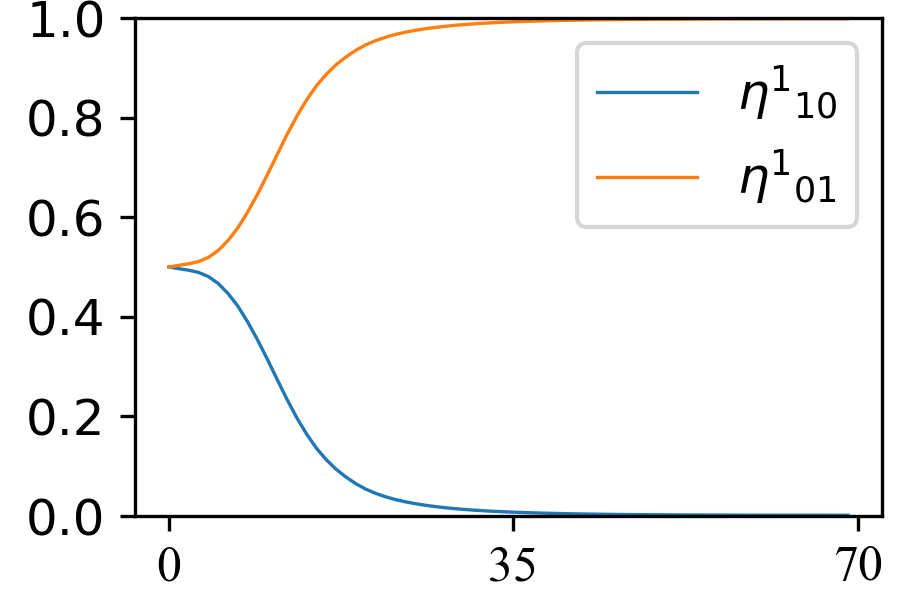}}
\subfigure[$\eta^2_{i,j},\;i+j=2$]{
    \label{fig:eta_Ada_2}
    \includegraphics[scale=0.54]{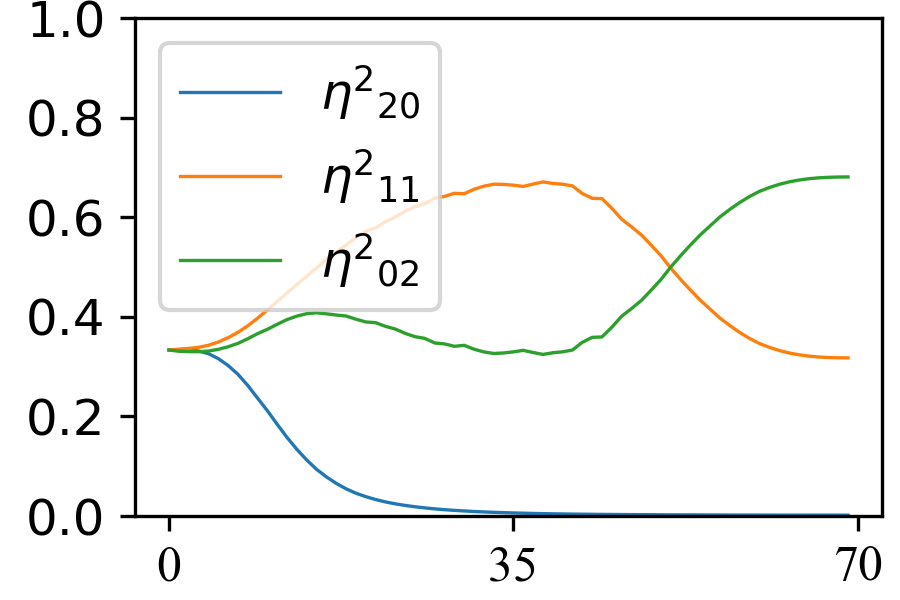}}
\subfigure[$\eta^3_{i,j},\;i+j=3$]{
    \label{fig:eta_Ada_3}
    \includegraphics[scale=0.54]{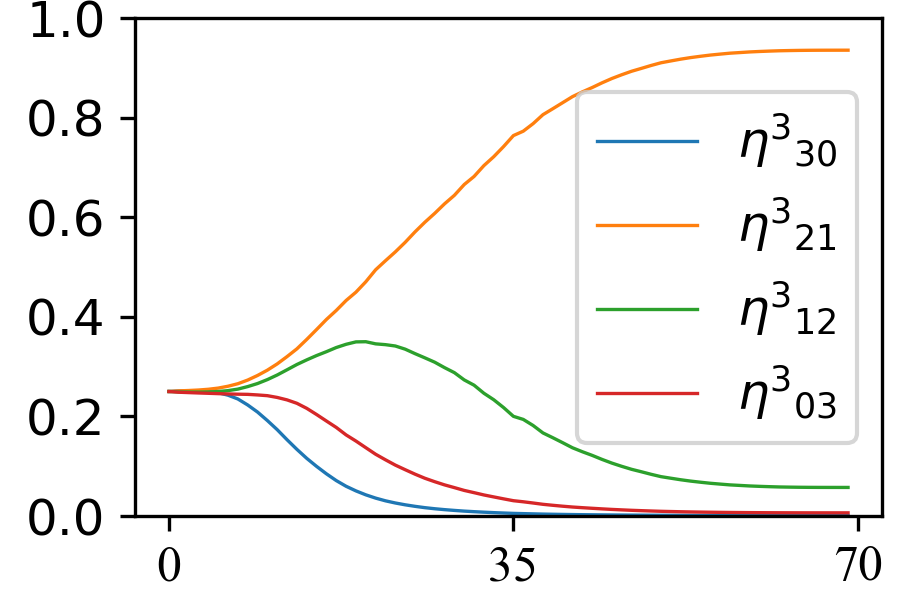}}
\subfigure[$\eta^4_{i,j},\;i+j=4$]{
    \label{fig:eta_Ada_4}
    \includegraphics[scale=0.54]{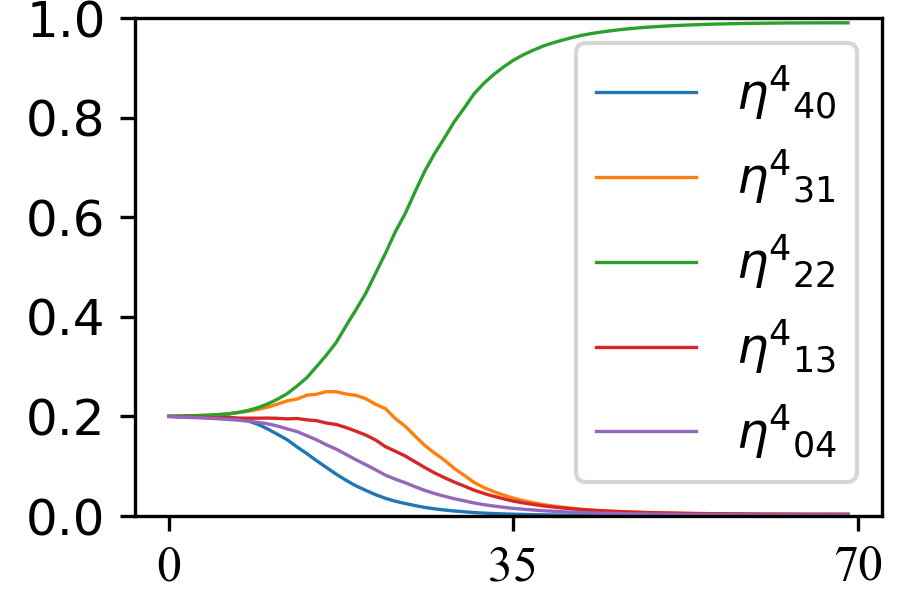}}
    \caption[The loss and the $\bm\eta^d$ of the UNet$^d$'s supervision layers.]{(a)-(d) Cross-entropy losses of supervision layers during the UNet$^d$ training process (Equation~\eqref{equation:constraint} is imposed to constraint the range of $\tilde{\eta}^d_{i,j}$). $\mathcal{L}_d$ is calculated from equation~\eqref{equ:overall-ce-loss_2}. (e)-(h) The corresponding weights of supervision layers. $\tilde{\eta}^d_{ij}$ reflects the importance of node X$^{i,j}$ while computing the overall loss. (i)-(p) The loss and the $\eta_{i,j}^d$ values of supervision layers of UNet$^d$, in which $\eta_{i,j}^d$ is trained without constraints. $\mathcal{L}_d$ shown in (i)-(l) is calculated from equation~\eqref{equ:overall-ce-loss}. For all plots, the x-axis indexes the training epoch. These plots are based on the BCSS dataset.}
	\label{fig:Ada_etas_BCSS}
\end{figure*}
\subsection{Deep Supervision in UNet$^\downarrow$ and ADS\_UNet}\label{section:analysis_1}
\subsubsection{Different layers contribute differently at different time stamps.} In UNet$^e$ and UNet++, all losses have the same weight in the back-propagation process, while in UNet$^\downarrow$ and ADS\_UNet, $\Tilde{\eta}_{i,j}$ is trainable. The purpose of this design is to check whether all layers in the summand of the training loss in equation~\eqref{equ:overall-ce-loss_2} contribute equally. Taking Figure ~\ref{fig:eta_DS_bcss} as an example, the importance of decoder nodes X$^{3,1}$ and X$^{2,2}$ is ranked in the top two. This means features learned by these 2 layers contribute more than others, with changes in their importance throughout the training process. From the perspective of back-propagation, this means that parameters of layers which have larger $\eta_{i,j}$ values, will have relatively large changes when they are updated using gradient descent. This fact, therefore, indicates the importance of the features derived at that length scale to the separability of texture labels. The segmentation performance achieved by layer $X^{3,1}$, $X^{2,2}$ and $X^{0,4}$(the last layer) are: 59.57\%, 59.55\% and 59.41\%, respectively. This is consistent with $\Tilde{\eta}_{i,j}$ values (see Figure ~\ref{fig:eta_DS_bcss}), $\eta_{3,1}>\eta_{2,2}>\eta_{0,4}$. A similar trend in the changes to $\Tilde{\eta}^d_{i,j}$ in the iterative training process of ADS\_UNet is also observed in Figure~\ref{fig:eta_Ada_const_1}-~\ref{fig:eta_Ada_const_4}.

Figure ~\ref{fig:eta_DS_bcss} and Figure ~\ref{fig:eta_Ada_const_1}-~\ref{fig:eta_Ada_const_4} not only show us how the parameters of different layers change during the training process, but also indicate that: 1) the importance of parameter varies from layer to layer; 2) the significance of parameters also vary throughout the training process. This is the effect of normalization of the weights $\Tilde{\eta}^d_{i,j}$, which introduces competition between the layers. And also, 3) the competition between the layers will continue until equilibrium is reached.
\begin{figure*}[!t]
    \label{fig:CKA_1}
    \centering
    \subfigure[UNet]{
      \label{fig:UNet_CKA}
      \includegraphics[scale=0.37]{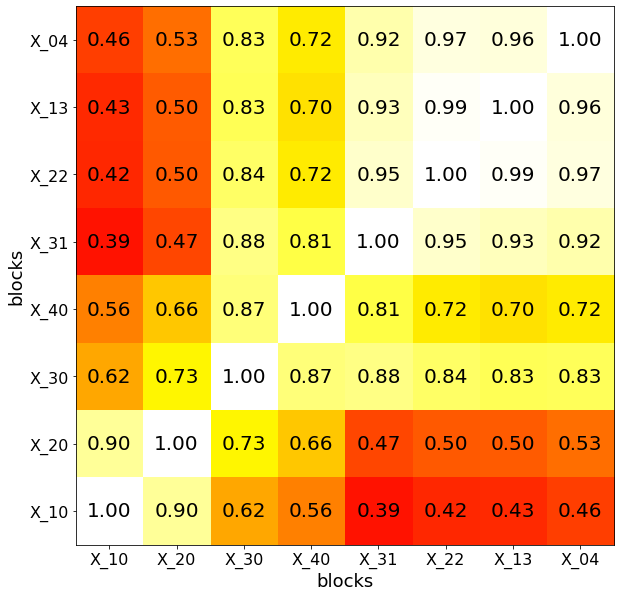}}
    \subfigure[diff(UNet$^\downarrow$, UNet)]{
      \label{fig:DS_UNet_CKA}
      \includegraphics[scale=0.37]{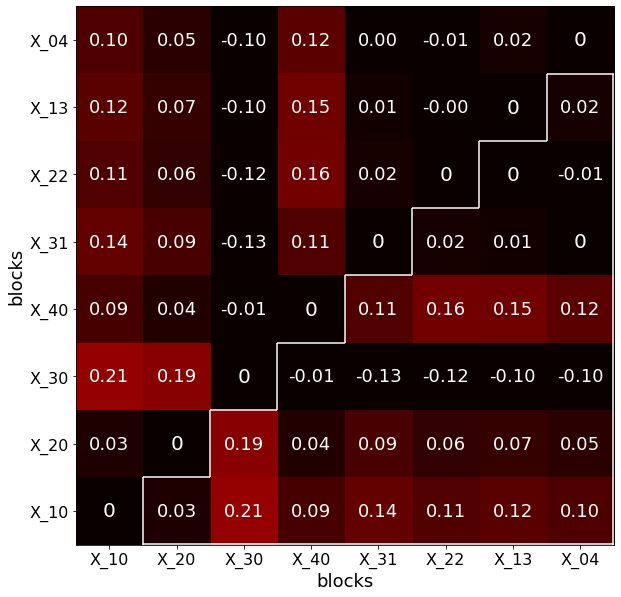}}
	\caption[CKA similarity of feature maps.]{(a) Feature similarity of layers for UNet. (b) The difference in feature similarity of layers between UNet$^\downarrow$ and UNet. In (a), each entry shows the CKA similarity between the two layers. In (b), we calculate the feature similarity matrix for UNet$^\downarrow$, then take the difference between UNet$^\downarrow$ and UNet. These plots are based on BCSS dataset.}
\end{figure*}
\begin{table}[!b]
	\centering
	\caption[mIoU score of ADS\_UNet trained in 3 modes]{mIoU score of ADS\_UNet trained in 3 modes based on the BCSS dataset. Each UNet$^d$ is trained for 70 epochs.} 
	\label{tab:training_criteria} 
	\begin{tabular}{lccccc}  
		\hline
		    & UNet$^1$ & UNet$^2$ & UNet$^3$ & UNet$^4$ & ens($\alpha$) \\
$\bm\eta^d$ & \textbf{47.77}   & 56.34 & 58.37 & 59.20 & 59.64 \\
$\bm{\tilde{\eta}}^d$ & 47.14  & 55.42 & 58.52 & 58.95 & 60.10 \\ 
$\bm{\tilde{\eta}}^d$(sum) & 46.93 & \textbf{56.91} & \textbf{60.11} & \textbf{60.26} & \textbf{61.05}\\\hline
	\end{tabular}
\end{table}
\subsubsection{Preventing $\eta^d_{i,j}$ from vanishing leads to higher segmentation performance.} In equation~\eqref{equation:constraint}, we redefine $\eta^d_{i,j}$ as $\tilde{\eta}^d_{i,j}$ to enforce all layers to learn features that are directly discriminative for classifying textures. We then sum the probability maps produced by these layers based on their importance factors $\tilde{\eta}_{i,j}^d$ to generate the segmentation map of UNet$^d$ (defined in equation~\eqref{equ:unetd_y_hat}). To verify if this constraint range and the weighted combination yield better performance or not, we run experiments on the BCSS dataset, in which ADS\_UNet is trained in 3 modes:
\begin{itemize}
    \item[1)] $\bm\eta^d$: with its element $\eta^d_{i,j}$ being trained without range constraint. After UNet$^d$ is trained, the output of the layer which has the largest $\eta_{i,j}^d$ value is selected to generate the final segmentation map. i.e., let $(i^\prime,j^\prime)={\arg\max}_{(i,j)}(\eta^d_{i,j})$, the final probability map is obtained by $\hat{y}^d=\hat{y}^{i^\prime,j^\prime}$, with $\hat{y}^{i,j}$ defined in equation~\eqref{equ:block_y_hat}. Then $\hat{y}^d$ is used to compare with the ground truth to calculate the $\alpha^d$ (the weight of UNet$^d$).
    \item[2)] $\bm{\tilde{\eta}}^d$: $\tilde{\eta}_{i,j}^d$ is bounded in $[\frac{1}{2(d+1)}\frac{d+2}{2(d+1)}]$, according to equation~\eqref{equation:constraint}. The final segmentation map generation and $\alpha^d$ calculation criteria are the same as 1).
    \item[3)] $\bm{\tilde{\eta}}^d$ (sum): training criteria is the same as 2). While the segmentation map produced by model UNet$^d$ is the weighted summation of multi-scale prediction (using equation~\eqref{equ:unetd_y_hat}), which is then used to calculate the $\alpha^d$.
\end{itemize}
The results of training ADS\_UNet in 3 different modes are reported in Table ~\ref{tab:training_criteria}, where ADS\_UNet with bounded $\bm{\tilde{\eta}}^d$ is seem to slightly surpass the unbounded one. After combining the probability maps produced by supervision layers based on the layer importance factors $\tilde{\eta}_{i,j}^d$, the mIoU score on the BCSS dataset is further improved by 0.95 points. To explain the results of Table ~\ref{tab:training_criteria}, the loss, $\eta^d_{i,j}$ and $\tilde{\eta}_{i,j}^d$ of the ADS\_UNet (trained in mode 1 and mode 3) are tracked and visualized in Figure ~\ref{fig:Ada_etas_BCSS}. As observed in Figure ~\ref{fig:loss_Ada_1}-\ref{fig:eta_Ada_4}, when there is no range constraint on $\eta^d_{i,j}$, only one specific layer's loss dominates the learning process and the loss of other layers is almost negligible ($\eta^d_{i,j}$ close to 0), after training for a few epochs. But the loss increases ($\mathcal{L}^{3,0}$ in Figure~\ref{fig:loss_Ada_3} and $\mathcal{L}^{4,0}$ in Figure~\ref{fig:loss_Ada_4}), so there is reduced discriminability at the intermediate layers (X$^{3,0}$, X$^{4,0}$) still. However, this phenomenon is eliminated after the range constraint is imposed, to suppress the weight of the dominant layer and to enable those of the others to grow, as shown in Figure ~\ref{fig:loss_Ada_const_1}-~\ref{fig:eta_Ada_const_4}. That means, by retaining the information from previous layers, the range of features that are being learned is increased, therefore leading to better performance. Note that $\mathcal{L}^{3,0}$ in Figure~\ref{fig:loss_Ada_const_3} and $\mathcal{L}^{4,0}$ in Figure~\ref{fig:loss_Ada_const_4} keep decreasing, and differs from that of Figure~\ref{fig:loss_Ada_3} and Figure~\ref{fig:loss_Ada_4}.
\subsection{Feature Similarity of Hidden Layers}\label{section:Feature_similarity}
\begin{figure*}[!t]
    \label{fig:CKA_2}
    \centering
    \subfigure[ADS\_UNet]{
        \label{CKA_ADS}
        \includegraphics[scale=0.34]{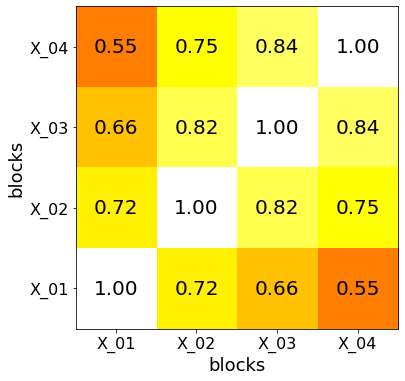}}
    \subfigure[diff(ADS\_UNet,UNet$^e$)]{
        \label{CKA_ADS-UNete}
        \includegraphics[scale=0.34]{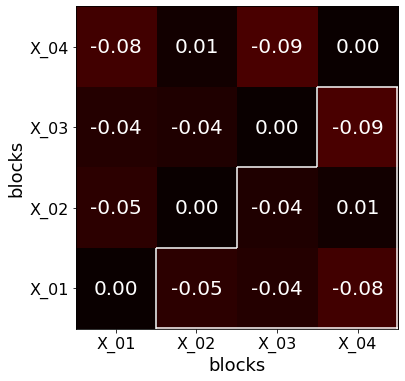}}
    \subfigure[diff(ADS\_UNet,UNet++)]{
        \label{CKA_ADS-UNet++}
        \includegraphics[scale=0.34]{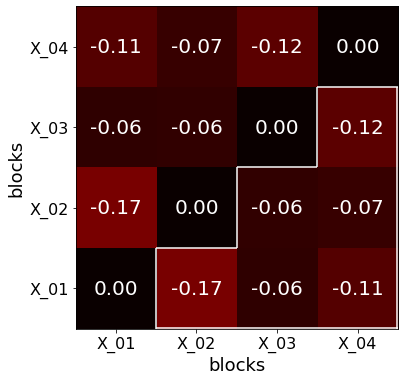}}
	\caption[CKA similarity of feature maps.]{(a) Feature similarity of the output layers of ADS\_UNet. (b) and (c) We calculate the feature similarity matrix for UNet$^e$ and UNet++, then take the difference between ADS\_UNet and UNet$^e$, UNet++.  These plots are based on BCSS dataset.}
\end{figure*}
Since deep supervision provides features of intermediate blocks with a direct contribution to the overall loss, the similarity of features learned by these blocks will be higher than those of the original UNet. Centered Kernel Alignment (CKA) \citep{kornblith2019similarity} has been developed as a tool for comparing feature representations of neural networks. Here we use CKA to characterize the similarity of feature representations learned by different convolutional blocks in UNet$^\downarrow$. As shown in Figure~\ref{fig:DS_UNet_CKA}, the similarity of features extracted by blocks in UNet$^\downarrow$ is mostly higher than in their counterparts in UNet (although 6 of similarity entries in UNet have lower values than that of UNet), which is consistent with our expectation (the 20 positive values add up to 1.89 vs. the 6 negative values add up to -0.47).
\subsection{Feature Diversity of Output Layers}\label{section:Feature_diversity}
Ensemble-based learning methods, such as AdaBoost, rely on the independence of features exploited by classifiers in its ensemble \citep{freund1997decision}. If base learners produce independent outputs, then the segmentation accuracy of the ensemble can be enhanced by majority weighting. Figure~\ref{CKA_ADS} characterize the feature similarity of output layers of ADS\_UNet. Figure~\ref{CKA_ADS-UNete} and ~\ref{CKA_ADS-UNet++} shows that features learned by the output layers of ADS\_UNet are less similar than those in UNet$^e$ (the values add up to -0.29) and UNet++ (the values add up to -0.59). Our interpretation is that this can be attributed to the stage-wise additive learning, followed by the sample weight updating rule of ADS\_UNet, and may explain why ADS\_UNet outperforms UNet$^e$ and UNet++.
\section{Conclusion}
In this paper, we propose a novel stage-wise additive training algorithm, ADS\_UNet, that incorporates the AdaBoost algorithm and greedy layer-wise training strategy into the iterative learning progress of an ensemble model. The proposed method has the following advantages: 1) The stage-wise training strategy with re-weighted training samples empowers base learners to learn discriminative and diverse feature representations. These are eventually combined in a performance-weighted manner to produce the final prediction, leading to higher accuracy than those achieved by other UNet-like architectures. 2) In the configuration of base learners, intermediate layers are supervised directly to learn discriminative features, without the need for learning extra up-sampling blocks. This, therefore, diminishes memory consumption and computational burden. 3) By introducing layer competition, we observe that the importance of feature maps produced by layers varies from epoch to epoch at the training stage, and different layers contribute differently in a manner that is learnable. 4) ADS\_UNet is more computationally efficient (fewer requirements on GPU memory and training time) than UNet$^e$, UNet++, CENet and transformer-based UNet variants, due to its cascade training regimen.

However, the ADS\_UNet has the following limitation that we would like to address in future work:  currently, the sample re-weighting training criteria restricts the ADS\_UNet to only update the weights of samples at a relatively coarse granularity. In future work, more fine-grained re-weighting criteria will be explored to guide successive base learners to pay more attention to regions/pixels that are difficult to distinguish. It would also be promising to integrate the AdaBoost and stage-wise training with a Transformer-like architecture to further improve segmentation performance.
\section*{Acknowledgement}
The authors acknowledge the use of the IRIDIS High-Performance Computing Facility, and associated support services at the University of Southampton, in the completion of this work. Yilong Yang is supported by China Scholarship Council under Grant No. 201906310150.
\bibliographystyle{cas-model2-names}

\bibliography{ESWA/ESWA}
\bio{}
\textbf{Yilong Yang} received the master degree in software engineering from Xiamen University, Xiamen, China, in 2019. He is currently a Ph.D candidate with Vision, Learning and Control Research Group, University of Southampton, United Kingdom. His research interests include computer vision and geometric deep learning.
\endbio

\bio{}
\textbf{Srinandan Dasmahapatra} received the Ph.D. degree in physics from the State University of New York, Stony Brook, NY, USA, in 1992. He is currently an Associate Professor in the School of Electronics and Computer Science, University of Southampton, Southampton, U.K. His research interests include artificial intelligence and pattern recognition.
\endbio

\bio{}
\textbf{Sasan Mahmoodi} received the Ph.D degree from the University of Newcastle, Newcastle upon Tyne, U.K., in 1998. He is currently an Associate Professor in the School of Electronics and Computer Science, University of Southampton, Southampton, U.K. His research interests include medical image processing, computer vision, and modeling of the biological vision.
\endbio

\appendix
\section{Python code for counting incorrect labels in down-scaled masks}\label{section:python}
\begin{lstlisting}
import numpy as np
def avg_pool(mask, mask_shape, window_size):
    err_count_per_mask = 0
    row, col = mask_shape/window_size
    for i in range(row):
        for j in range(col):
            window = img[i:i+window_size,j:j+window_size]
            if len(np.unique(window))>1:
                err_count_per_mask+=1
    return err_count_per_mask
def get_incorrect_ratio(mask_shape, mask_list):
    down_factor = [2, 4, 8, 16]
    index       = [0, 1, 2, 3]
    err_count_per_scale = [0, 0, 0, 0]
    for mask in mask_list:
        for idx,f in zip(index, down_factor):
            err_count_per_scale[idx] += avg_pool(mask, mask_shape, f)
    pixel_count = [mask_shape/f for f in down_factor]**2*len(mask_list)
    return np.array(err_count_per_scale)/np.array(pixel_count)
\end{lstlisting}
\end{document}